\documentclass[trackchanges]{aastex701}
\usepackage{color}
\usepackage{makecell}
\usepackage{threeparttable}

\begin{document}

\title{Investigating the Effects of Bars on Star Formation and Nuclear Activity of Galaxies Using DESI Survey Data}

\author{Jianfei Liu}
\affil{National Astronomical Observatories, Chinese Academy of Sciences, Beijing 100101, People's Republic of China}
\affil{School of Astronomy and Space Science, University of Chinese Academy of Sciences, Beijing 100049, People's Republic of China}
\email[]{liujf@nao.cas.cn}

\author[0000-0002-4135-0977]{Zhimin Zhou}
\affil{National Astronomical Observatories, Chinese Academy of Sciences, Beijing 100101, People's Republic of China}
\email[show]{zmzhou@nao.cas.cn}
\makeatletter\def\Hy@Warning#1{}\makeatother 
\correspondingauthor{Zhimin Zhou}

\begin{abstract}

We present a statistical analysis of the connections between galactic bars, star formation, and active galactic nucleus (AGN) activity using 33,201 disk galaxies ($0.01 < z < 0.05$) from DESI DR1 cross-matched with Galaxy Zoo DESI. Based on morphological classifications, we identify 3,508 strongly barred and 8,335 weakly barred systems.
We find that barred galaxies exhibit a clear bimodal distribution in color-mass space: weak bars are preferentially found in bluer, lower-mass disks, whereas strong bars are more common in massive, redder systems. Strongly barred galaxies are on average more massive and metal-rich than unbarred systems. In addition, strong bars enhance central SFRs in low-mass galaxies but reduce sSFRs in massive systems, reflecting a dual role where bars initially trigger central star formation but eventually promote quenching by accelerating gas consumption.
In terms of nuclear activity, barred galaxies display a higher incidence of AGN activity. The presence of a bar is also associated with an increased fraction of powerful AGN, with the highest proportions found in strongly barred systems. However, the correlations between AGN activity and detailed bar structural parameters are weak, suggesting that the link between bars and nuclear activity is indirect and regulated by multiple factors. 
Overall, our results support a scenario in which bars facilitate angular-momentum transport and gas inflow, thereby driving central star formation and fueling supermassive black hole accretion while operating alongside other processes that shape galaxy evolution.

\end{abstract}


\section{Introduction}

Galactic bars are ubiquitous non-axisymmetric structures in disk galaxies, and are found in a substantial fraction of spiral systems. While the observed bar fraction depends on wavelength and methodology, studies of the local Universe consistently find bars in one-third to half of disk galaxies at optical wavelengths \citep[][]{2008ApJ...675.1194B(bf0.5), 2021MNRAS.507.4389G}, and the fraction is even higher in infrared wavelengths, reaching $\sim70\%$ or more \citep[][]{2000AJ....119..536E, 2007ApJ...657..790M}, due to the reduced impact of dust obscuration and the improved visibility of old stellar bars. At higher redshift, however, the bar fraction exhibits a steady decline but remains non-negligible: at $z\sim1$ it is about $\sim20\%$ \citep[][]{2025arXiv250315311E(bf0.2-0.4)}, at $1<z<2$ the fraction drops to $\sim17.8\%$ \citep[][]{2024MNRAS.530.1984L(bf0.17_z3)}, and at $2<z<4$ it reaches $10\%$ \citep[][]{2025ApJ...985..181G(bf0.1_z4)}.

Conventionally, bars are commonly divided into `strong' and `weak' classes, although the criteria for this division are not always consistent across studies. Traditional approaches often rely on visual inspection, distinguishing bars by how evident or extended they appear \citep[][]{1959HDP....53..275D}, while some works classify bars based on whether the bar component dominates the galaxy’s light \citep[][]{2010ApJS..186..427N}. To provide more objective metrics, quantitative methods have also been employed, such as using bar ellipticity as a proxy for bar strength \citep[][]{1999A&A...351...43A, 1997A&A...323..363M}, while the ratio between the amplitudes of the Fourier m = 2 and m = 0 components is also used to quantify bar intensity\citep[][]{2000A&A...361..841A}. Whether this classification represents two distinct populations remains a subject of debate. Some studies suggest that bar strength is a continuous spectrum with no fundamental difference between strong and weak bars \citep[][]{2019A&A...632A..51C, 2021MNRAS.507.4389G}, while others propose that they may have distinct formation mechanisms \citep[][]{2023MNRAS.521.1775G}.

Galaxy Zoo DESI \citep[GZD,][]{2023MNRAS.526.4768W} provides large-scale morphological classifications based on crowdsourced visual inspection, enabling robust classification for both strong and weak bars. GZD is built on the deep, homogeneous $g,r,z$ imaging from the Dark Energy Spectroscopic Instrument (DESI) Legacy Imaging Surveys \citep[DESI-LS,][]{2019AJ....157..168D(DESILS)} and performs morphological classification through a combination of citizen-science visual inspection and a structured decision-tree workflow \citep{2023MNRAS.526.4768W}. 
Compared with earlier studies based on shallower imaging or purely visual classifications, GZD benefits from deeper and more uniform data, a much larger sample size, probabilistic morphology indicators, and the complementary strengths of human annotation and machine-learning inference. These advantages make GZD particularly effective at identifying weak or low surface brightness bars, providing a robust basis for statistical studies of correlations between bars and galaxy properties.

Bars are recognized as key drivers of secular evolution, efficiently redistributing gas, stars and angular momentum within the galactic disk \citep[][]{1989Natur.338...45S, 1999ApJ...525..691S, 2004ARA&A..42..603K}. By exerting gravitational torques, bars funnel large quantities of gas from the outer disk into the inner kiloparsec, often triggering intense central star formation \citep[][]{2005ApJ...632..217S, 2011MNRAS.416.2182E, 2012MNRAS.423.3486W, 2020MNRAS.499.1406L}. Beyond gas dynamics, bars may radially mix stars, flattening stellar age and metallicity profiles \citep[][]{2019MNRAS.488L...6F}. Over longer timescales, the gas reservoir may be depleted by these bar-driven inflows, ultimately leading to galaxy quenching \citep[][]{2021A&A...651A.107G, 2020MNRAS.499.1116F, 2021MNRAS.507.4389G}. 

The same inflow that fuels central star formation is also believed to feed the central supermassive black hole (SMBH), thereby connecting with active galactic nucleus (AGN). Many theoretical studies have explored this connection, ranging from high-resolution idealized simulations to cosmological frameworks. Recent numerical evidence suggests that gas transport is not a steady process but is significantly enhanced by the dynamic bar-formation phase \citep[][]{2015MNRAS.454.3641F} and modulated by multi-scale structures such as nuclear rings and spirals in mature barred systems \citep[][]{2013ApJ...771....8L}. Furthermore, large-scale cosmological simulations like TNG50 have statistically corroborated a robust correlation between the presence of strong bars and increased central black hole accretion rates across cosmic time \citep[][]{2022MNRAS.512.5339R}.

Observationally, this idea has also been investigated extensively, however, whether bars are associated with AGN activity remains unclear. Some surveys indeed find higher AGN fraction or enhanced AGN signatures in barred disks, suggesting a direct causal link. \citep[][]{2002ApJ...567...97L, 2012ApJS..198....4O, 2018A&A...618A.149A, 2023MNRAS.522..211G, 2024MNRAS.532.2320G, 2025A&A...699A.204M}. Conversely, other studies report little or no significant difference between barred and unbarred AGN hosts, proposing instead that the observed associations may be driven by underlying host-galaxy properties, rather than the bar structure itself. \citep[][]{2012ApJ...750..141L, 2015MNRAS.447..506C, 2017ApJ...843..135G(nolink)}. For instance, while \citet{2015MNRAS.448.3442G} find higher bar fraction in AGN host galaxies, but also indicated that barred AGNs do not exhibit stronger accretion than unbarred AGNs at fixed stellar mass and color. These conflicting results may stem from sample selection, AGN diagnostics, or the range of galaxy parameters probed, underscoring the need for large, statistically robust samples to disentangle the complex interplay between galactic bars and nuclear activity.

In this work, we leverage the large, morphologically-classified Dark Energy Spectroscopic Instrument Data Release 1 \citep[DESI DR1,][]{2026AJ....171..285D} galaxy sample to systematically revisit the connections between bars, host galaxy properties, and nuclear activity. By utilizing morphologies from the Galaxy Zoo DESI project, we ensure a robust and consistent identification of barred structures across a large, representative sample of the local Universe. The paper is organized as follows. Section~\ref{sec:data} describes sample selection and the value-added catalogs used. Section~\ref{sec:results} presents the results of our analysis, focusing on the observed distributions and trends. Section~\ref{sec:discussion} discusses the implications for secular evolution and AGN fueling. Finally, Section~\ref{sec:summary} summaries our main results and conclusions.

Throughout this work, we adopt the WMAP7 cosmology \citep{2011ApJS..192...18K}, with $H_0 = 70.4,\mathrm{km,s^{-1},Mpc^{-1}}$, $\Omega_m = 0.272$, and $\Omega_\Lambda = 0.728$.


\section{Data and Sample}
\label{sec:data}
In this section, we detail the construction of our primary galaxy sample, integrated from the DESI DR1 and Galaxy Zoo DESI datasets. Then we outline the Value-Added Catalogs used to obtain derived physical parameters, which together form the framework for our analysis.

\subsection{Sample selection}
\label{subsec:sample_selection}
The galaxy sample analyzed in this study is based on data from the DESI DR1. 
DESI is a spectroscopic survey designed to obtain optical spectra for tens of millions of galaxies and quasars 
over an area of about 14,000~deg$^2$, providing redshifts and spectral classifications for studies of large-scale structure and galaxy evolution. 

To obtain reliable morphological classifications, the DESI DR1 catalog was cross-matched with the GZD catalog. 
GZD is a citizen science project that provides visual morphological classifications of DESI imaging data, offering probabilistic measures for features such as bars, disks, and spiral structures. Cross-matching was performed using right ascension (RA) and declination (DEC) coordinates with a matching radius of $3^{\prime\prime}$, ensuring accurate associations between the two catalogs while minimizing duplicate matches or stellar contamination.

From the cross-matched catalog, galaxies were selected within the redshift range $0.01 < z < 0.05$ and with absolute r-band magnitudes brighter than $-17.25$~mag. The photometry is taken from the DESI Legacy imaging Surveys. The redshift limit ensures that galaxies are sufficiently nearby for reliable morphological classification while maintaining a statistically meaningful sample size. Furthermore, the magnitude cut is implemented to guarantee volume-completeness across the entire adopted redshift range, providing a statistically representative population for our analysis.

Disk galaxies were identified by applying morphological probability thresholds following the classification scheme suggested by \citet{2021MNRAS.507.4389G}:
\begin{equation}
(p_{\text{featured-or-disk}} \geq 0.27) 
\ \&\ 
(p_{\text{disk-edge-on\_no}} \geq 0.68)
\end{equation}
Here $p_{\text{featured-or-disk}}$ denotes the vote fraction that a galaxy exhibits visible features or a disk component, while $p_{\text{disk-edge-on\_no}}$ represents the vote fraction that the galaxy is not viewed edge-on. Galaxies satisfying these criteria were classified as disk galaxies. 
Both the full galaxy sample and the disk-galaxy subsample were considered separately in the subsequent analysis.

Barred and unbarred galaxies were distinguished using the parameter \( p_{\text{bar-no}} \), which represents the vote fraction that no bar is identified in a galaxy. Adopting a threshold of 0.5, galaxies with \( p_{\text{bar-no}} < 0.5 \) were identified as barred, while those with higher values were classified as unbarred. 
The barred population was further divided into strongly barred and weakly barred systems according to the relative magnitudes of \( p_{\text{bar-strong}} \) and \( p_{\text{bar-weak}} \), which give the vote fractions for strong and weak bars, respectively.

From the parent sample of 99,998 galaxies, we identify 33,201 disk galaxies based on the adopted criteria. The final sample statistics are summarized as follows:

\begin{itemize}
    \item Disk-galaxy subsample: 33,201 galaxies
    \begin{itemize}
        \item Strong bars: 10.57\% (3,508/33,201)
        \item Weak bars: 25.10\% (8,335/33,201)
        \item Barred: 35.67\% (11,843/33,201)
        \item Unbarred: 64.33\% (21,358/33,201)
    \end{itemize}
\end{itemize}

The resulting dataset provides a large and morphologically well-characterized sample 
of nearby galaxies suitable for statistical analysis of bar occurrence and strength.

\subsection{Value-Added catalogs and derived parameters}
\label{subsec:vac_parameters}

In this study, we used several value-added catalogs (VACs) from the DESI DR1 to obtain reliable measurements of galaxy physical and spectroscopic properties. 
Specifically, these catalogs provide derived  total and fiber-based quantities such as stellar mass, star formation rate (SFR), emission-line fluxes, and spectral indices, based on homogeneous analyses of DESI DR1 spectra and imaging data. 
The following VACs were used in this work:

\begin{enumerate}

\item \textbf{AGN Host Galaxies Physical Properties VAC:}  
This catalog provides physical parameters for approximately 17 million DESI DR1 galaxies, 
derived through spectral energy distribution (SED) fitting with the \textit{Code Investigating GALaxy Emission} (CIGALE v.22.1; \citealt{2019A&A...622A.103B}), which self-consistently accounts for the contribution of AGN. CIGALE is based on the principle of energy balance between dust-absorbed stellar emission in the ultraviolet and optical bands and its re-emission in the infrared, and has been widely applied in galaxy surveys \citep[][]{2015A&A...576A..10C, 2018A&A...620A..50M, 2018ApJ...859...11S, 2021A&A...646A..29M, 2020MNRAS.491..740Y, 2022ApJ...927..192Y}. The VAC construction and statistical validation are described in detail in \citet{2024A&A...691A.308S}.

The SED fitting grid adopts a delayed star formation history (SFH) model with an optional exponential burst and the \citet{2003MNRAS.344.1000B} single stellar population templates, assuming a \citet{2003PASP..115..763C} initial mass function and solar metallicity. Nebular emission is modeled following \citet{2011MNRAS.415.2920I}, dust attenuation follows the \citet{2000ApJ...533..682C} law, and dust re-emission adopts the models of \citet{2014ApJ...784...83D}. The AGN component is modeled using the templates from \citet{2006MNRAS.366..767F}, allowing simultaneous fitting of AGN and stellar contributions across the available photometric bands ($g$, $r$, $z$, W1–W4). The resulting physical parameters, such as stellar mass, SFR, and AGN fraction, are computed as likelihood-weighted means of the posterior probability distributions, with uncertainties corresponding to their standard deviations.

In this study, we characterize the global properties of our galaxy sample using the logarithm of the stellar mass (\(\log M_\ast\)), the logarithm of SFR averaged over 10~Myr (\(\log \mathrm{SFR}\)), and the extinction-corrected $(g-r)$ color retrieved from this catalog.

\item \textbf{FastSpecFit Spectral Synthesis and Emission-Line Catalog:}  
The \textit{FastSpecFit} catalog is generated using a stellar continuum and emission-line modeling code 
developed for DESI spectroscopy, optimized for speed and physical interpretability. 
It jointly fits the DESI three-camera optical spectra together with ultraviolet–to–infrared broadband photometry, 
employing stellar population synthesis and emission-line templates to extract key physical parameters of galaxies.  

In this study, we adopt some several parameters from the FastSpecFit catalog, including the 4000~\AA\ break index ($D_n4000$) and the fiber-based SFR. The former traces the strength of the 4000~\AA\ break, a widely used diagnostic of stellar population age, while the latter quantifies the star formation rate measured within the spectroscopic fiber aperture, providing an estimate of the local, central star forming activity. We also make use of the fiber and total galaxy luminosities to estimate the stellar mass within the fiber region. 
These quantities enable us to examine the relationship between bar structures, stellar age, and central star formation in galaxies.

\item \textbf{AGN/Galaxy Classification VAC:}
This catalog includes DESI DR1 galaxies and quasars from all primary and secondary target classes 
(MWS, BGS, LRG, ELG, and QSO; \citealt{2023AJ....165...50M}). 
The DESI Redrock pipeline assigns a spectral type (\texttt{SPECTYPE} = \texttt{‘STAR’}, \texttt{‘GALAXY’}, or \texttt{‘QSO’}) 
and a redshift to each object, while the machine-learning algorithm QuasarNet 
and the [Mg II] post-processing pipeline provide refined classifications and redshift estimates 
(\citealt{2023ApJ...944..107C}; \citealt{2023AJ....165..124A}). 
The catalog further integrates optical and ultraviolet diagnostics based on emission-line measurements from 
\texttt{FastSpecFit} (\citealt{2023ascl.soft08005M}) 
with mid-infrared AGN classifications derived from Tractor WISE photometry (\citealt{2016ascl.soft04008L}). 

The AGN/Galaxy Classification VAC provides emission-line excitation classifications based on the Baldwin–Phillips–Terlevich (BPT; \citealt{BPT1981PASP...93....5B}) diagrams constructed using the [N II]/H$\alpha$, [S II]/H$\alpha$, and [O I]/H$\alpha$ intensity ratios relative to [O III]/H$\beta$ \citep[][]{BPT1981PASP...93....5B, 1987ApJS...63..295V}. These diagnostics adopt commonly used demarcation schemes \citep[e.g.,][]{2001ApJ...556..121K, 2003MNRAS.346.1055K, 2007MNRAS.382.1415S} to distinguish star-forming galaxies from composite systems and AGN (including Seyferts and LINERs). In this work, we identify star-forming galaxies using the classification from the [N II] based BPT diagram provided in the \texttt{OPT\_UV\_TYPE} parameter, requiring that the diagnostic is available and that the galaxy is flagged as lying in the star-forming region. This choice is motivated by methodological consistency, since the O3N2 gas-phase metallicity calibration we used in this work is defined using the same set of emission lines involved in the corresponding [N II] BPT diagnostic.

Besides, we use the \texttt{AGN\_MASKBITS} parameter from the AGN/Galaxy Classification VAC, which encodes AGN selection flags based on multiple diagnostic methods, including those for which at least one BPT diagram classifies them as \textsc{Seyfert}.
These classes correspond to galaxies exhibiting strong or moderate nuclear ionization from accreting supermassive black holes, 
distinguishable from purely star-forming systems through their characteristic emission-line ratios.

\item \textbf{EMFit Catalog:}  
The \textit{EmFit} catalog is produced using the Python-based emission-line fitting code \textit{EmFit} \citep{2025ApJ...982...10P}, which performs simultaneous multi-component fitting of the H$\beta$, [O III] $\lambda4959,\lambda5007$, [N II] $\lambda6548,\lambda6583$, H$\alpha$, and [S II] $\lambda6716,\lambda6731$ emission lines for low-redshift ($z \leq 0.45$) galaxies. The code also tests for the presence of additional kinematic components in the [O III], H$\alpha$, and [S II] lines, allowing the identification of outflowing gas or multiple ionized regions.  

In this work, we extracted the [O III] $\lambda5007$ luminosity ($L_{\text{[O III]}}$) from the EmFit catalog. The [O III] $\lambda5007$ line is a key tracer of highly ionized gas and is commonly used to diagnose AGN activity, as it originates in the narrow line region (NLR) photoionized by the AGN’s central radiation field. The strength of this line correlates with the AGN ionizing luminosity, providing a proxy for the AGN’s power and contribution to the host galaxy’s emission. Using the [O III] $\lambda5007$, [N II] $\lambda6584$, H$\alpha$, and H$\beta$ line fluxes, we derived the gas-phase metallicity employing the O3N2 diagnostic \citep[][]{2004MNRAS.348L..59P(o3n2)}, defined as:
\begin{equation}
\mathrm{O3N2} \equiv 
\log_{10}\left(
\frac{[\mathrm{O\,III}]\lambda5007 / \mathrm{H}\beta}
{[\mathrm{N\,II}]\lambda6584 / \mathrm{H}\alpha}
\right).
\end{equation}
Following the empirical calibration established by \citet{2013A&A...559A.114M(o3n2)}, the oxygen abundance is calculated as:
\begin{equation}
12+\log(\mathrm{O/H}) = 8.533 - 0.214 \times\mathrm{O3N2}.
\end{equation}
In this work, gas-phase metallicities are computed only for galaxies classified as star-forming in the AGN/Galaxy Classification VAC based on the [N II] based BPT diagnostic. This selection ensures that the adopted O3N2 calibration, which is empirically derived for H II regions, is applied within its valid regime. The metallicity is expressed as a relative abundance, $Z/Z_\odot$, computed by normalizing $12+\log(\mathrm{O/H})$ to the solar oxygen abundance of $12+\log(\mathrm{O/H})_\odot = 8.69$.
\end{enumerate}
These VACs collectively provide a comprehensive set of photometric and spectroscopic indicators, 
allowing for a consistent and multi-dimensional characterization of the galaxy sample.

\subsection{Bar Size and Ellipticity Measurements}

The bar structural parameters used in this work, including deprojected bar length, normalized bar length, and bar ellipticity, are taken from the catalog of \citet{2025ApJ...982..129W}. Their measurements are based on the classical \texttt{IRAF} \texttt{ELLIPSE} task, which fits a sequence of elliptical isophotes to the DESI Legacy Survey images to extract radial profiles of surface brightness, ellipticity, and position angle.

The observed bar radius is defined as the semimajor axis at the peak ellipticity ($e_{\rm max}$) of the radial ellipticity profile, and the deprojected bar length $R_{\mathrm{bar}}$ is computed following the geometry-correction procedure described by \citet{2011ApJS..197...22L} to remove line-of-sight projection effects using the disk axis ratio and position angles. The normalized $R_{\mathrm{bar}}$ is obtained by dividing the deprojected bar length by the disk size $R_{25}$ at the $r$-band 25~mag~arcsec$^{-2}$ isophote. The bar ellipticity ($e_\mathrm{bar}$) is measured at the same radius and defined as $e = 1 - b/a$, with $b/a$ corrected to the face-on plane.
Benefiting from the large, well-resolved DESI Legacy Survey sample and the catalog’s uniform methodology, the structural parameters used in this work are robust and suitable for statistical analysis. The subsample with available bar structural parameter measurements contains 7,425 barred disks, including 2,427 strongly barred and 4,998 weakly barred systems.

   \begin{figure*}
        \centering
        \includegraphics[width=0.93\textwidth]{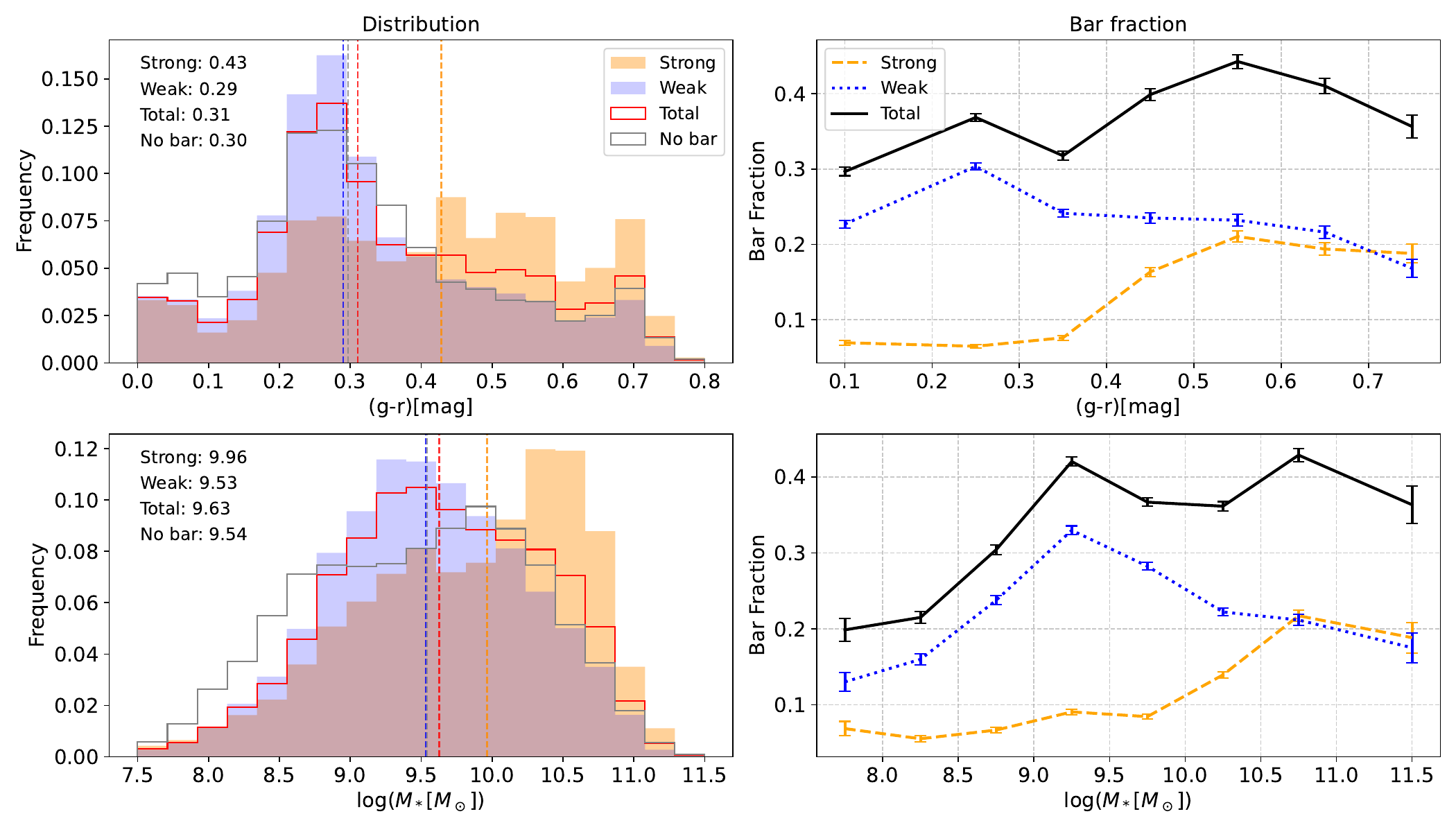}
        \caption{The color and stellar mass characteristics of the disk sample. The left panels show the distributions of $(g-r)$ and $M_{\ast}$ of barred and unbarred galaxies. All frequencies are normalized. Strongly barred galaxies are shown by orange filled histograms, weakly barred galaxies are shown by blue filled histograms. The total barred sample is indicated by red open histograms, and unbarred galaxies are indicated by gray open histograms. Dashed lines in orange, blue, red, and gray mark the medians of the four subsamples. Median values are shown in the upper-left corner of each panel. The right panels show the dependence of the bar fraction on $(g-r)$ and $M_{\ast}$. Orange dashed, blue dotted, and black solid lines represent strong, weak, and total barred galaxies, respectively. Error bars indicate the binomial uncertainties of the bar fraction in each bin.
}
        \label{GR_M}%
    \end{figure*}

\begin{table}
\centering
\begin{threeparttable}
\caption{K-S test results for different subsamples.}
\label{tab:ks}
\begin{tabular}{lcccc}
\hline
Parameter & Strong vs. Weak & Strong vs. No Bar & Weak vs. No Bar & Total vs. No Bar \\
\hline
$g-r$ & 0.27 $(p < 0.001)$ & 0.27 $(p < 0.001)$ & 0.04 $(p < 0.001)$ & 0.07 $(p < 0.001)$ \\
$M_*$ & 0.22 $(p < 0.001)$ & 0.20 $(p < 0.001)$ & 0.09 $(p < 0.001)$ & 0.10 $(p < 0.001)$ \\
$\log(\mathrm{SFR})$ & 0.09 $(p < 0.001)$ & 0.10 $(p < 0.001)$ & 0.02$(p = 0.200)$ & 0.04 $(p < 0.001)$ \\
$\log(\mathrm{fiber\,SFR})$ & 0.07 $(p < 0.001)$ & 0.05 $(p < 0.001)$ & 0.02$(p = 0.210)$ & 0.02$(p = 0.190)$ \\
$D_n4000$ & 0.11 $(p < 0.001)$ & 0.07 $(p < 0.001)$ & 0.08 $(p < 0.001)$ & 0.06 $(p < 0.001)$ \\
$\log(Z/Z_{\odot})$ & 0.11 $(p < 0.001)$ & 0.08 $(p < 0.001)$ & 0.07 $(p < 0.001)$ & 0.06 $(p < 0.001)$ \\
\hline
\end{tabular}
\begin{tablenotes}
\footnotesize
\item Note: For the $g-r$ color and stellar mass $M_\ast$, the `No Bar' sample refers to the original unbarred disk sample. For other parameters, the `No Bar' sample corresponds to the control sample constructed to match the barred population in stellar mass and color.
\end{tablenotes}
\end{threeparttable}
\end{table}

   \begin{figure*}
        \centering
        \includegraphics[width=1.\textwidth]{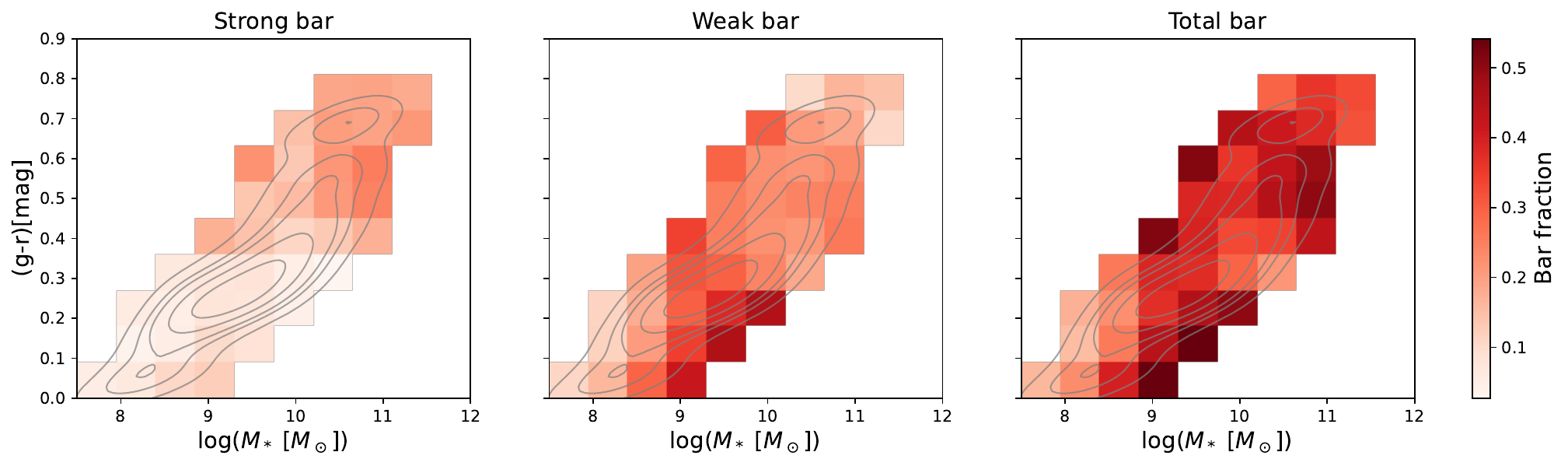}
        \caption{Bar fraction in the $(g - r)$–$M_\ast$ plane for the disk galaxy sample. The left, middle, and right panels show the fractions for strong bar, weak bar, and the total bar samples, respectively. All bins are uniformly divided in both $(g - r)$ and ($M_\ast$), The bar fraction in each bin is indicated by the color intensity. Gray contours represent the distribution of all disk galaxies in the sample.
}
        \label{frac_gr_M}%
    \end{figure*}

\section{Analysis and Results}
\label{sec:results}

We examined several key galaxy parameters for disk galaxies, including global properties and fiber-based spectral properties. 
The global properties are $(g–r)$ color, stellar mass ($M_\ast$), and total SFR. We also analyzed central, fiber-based parameters from the spectra, including fiber SFR, $D_n4000$, and $Z/Z_{\odot}$.

\subsection{Stellar Mass and Color Distributions}
\label{subsec:M and (g-r)}

The distributions of $(g-r)$ and $M_{\ast}$ were analyzed separately for strongly barred, weakly barred, total barred, and unbarred galaxies, as shown in the left panels of Figure~\ref{GR_M}, while the right panels show the dependence of the bar fraction on $(g-r)$ color and $M_{\ast}$.
The Kolmogorov Smirnov test \citep[KS test,][]{kolmogorov_1951} was performed between all subsamples, with the null hypothesis that the two samples are drawn from the same parent distribution, and the results indicate a statistically significant difference between the two distributions. The corresponding KS statistics and $p$-values are presented in Table~\ref{tab:ks}

The $(g - r)$ color distribution reveals distinct trends across the three populations. Weakly barred and unbarred galaxies show a prominent peak around $(g - r) \approx 0.3$, 
while strongly barred galaxies are more uniformly distributed over the range $(g - r) = 0.2$–$0.8$.  
All subsets exhibit a noticeable depletion near $(g - r) \sim 0.6$, corresponding to the so-called “green valley.”  
Examining the bar fraction as a function of color, we find that weak bar fraction is higher at the blue end. Conversely, the strong bar fraction increases significantly at $(g-r) > 0.3$, with the difference between strong and weak bar incidences diminishing toward redder colors. This suggests that weak bars tend to occur in bluer, star-forming galaxies while strong bars are more frequently found in more evolved, redder galaxies.

From the $M_{\ast}$ distribution, barred galaxies are preferentially found in more massive systems, showing relatively enhanced normalized frequencies among galaxies with $M_\ast > 10^{10.3}M_{\odot}$ compared with unbarred systems, whereas unbarred galaxies exhibiting higher normalized frequencies across galaxies with $M_\ast < 10^{9}M_{\odot}$. Strongly barred galaxies show a clear peak at $M_\ast \sim 10^{10.5} M_{\odot}$, whereas weakly barred galaxies peak at lower masses, around $M_\ast \sim 10^{9.5} M_{\odot}$. The total barred sample exhibits elevated frequencies near both peaks compared to the unbarred sample. The bar fraction shows a bimodal pattern on stellar mass. The fraction of weak bars reaches a maximum near $M_\ast \sim 10^{9.5} M_{\odot}$ and declines toward higher masses. In contrast, the strong bar fraction increases with stellar mass and peaks around $M_\ast \sim 10^{10.5} M_{\odot}$, becoming comparable to weak bar fraction.

We further examined how the bar fraction varies jointly with color and stellar mass by mapping the fractions of strong, weak, and total bars in the $(g - r)$--$M_\ast$ plane for the disk galaxy sample, as shown in Figure~\ref{frac_gr_M}. The plane was divided into uniform bins in both color and stellar mass, and the color intensity of each cell represents the corresponding bar fraction. To ensure statistical robustness, we only consider bins containing at least 50 galaxies. Gray contours mark the regions enclosing 95\%, 80\%, 60\%, 40\%, and 20\% of the total galaxy population, providing a visual reference for the overall distribution. The distributions show that strongly barred galaxies are primarily concentrated in intermediate-mass systems ($10 < \log(M_\ast/M_\odot) < 11$) with moderately red colors ($0.4 < g - r < 0.6$), while weakly barred galaxies are more common in lower-mass ($9 < \log(M_\ast/M_\odot) < 10$), bluer ($g - r < 0.3$) systems. As a result, the total bar fraction exhibits a bimodal structure across the $(g - r)$--$M_\ast$ plane.

\subsection{Bar Length and Ellipticity Distributions}
\label{subsubsec:bar_length_ellipticity}

\begin{figure*}
        \centering
        \includegraphics[width=\textwidth]{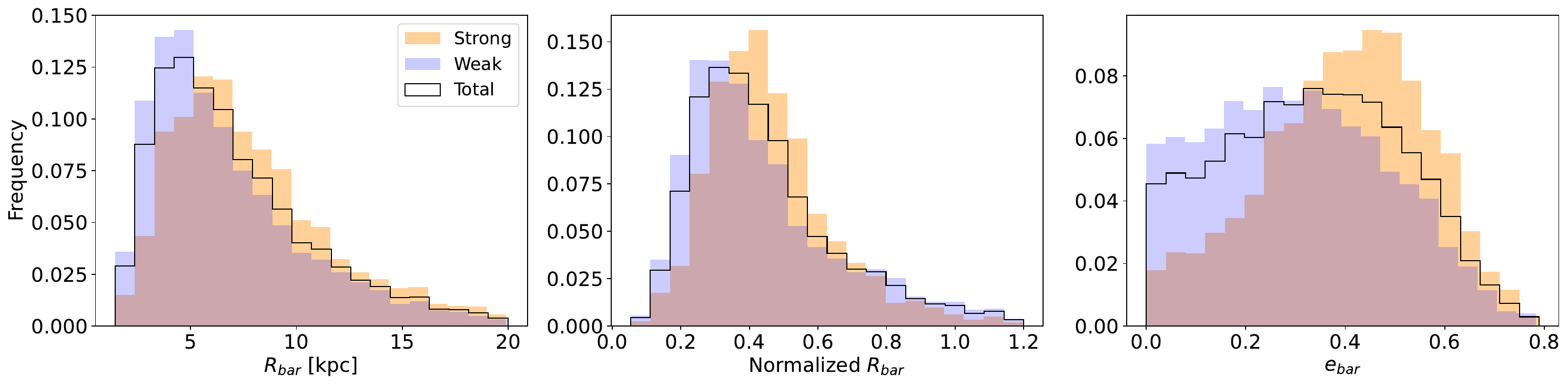}
        \caption{Distributions of bar length $R_{bar}$, normalized $R_{bar}$, and bar ellipticity $e_{bar}$ for the disk barred sample. The strong and weak bar subsamples are shown as orange and blue filled histograms, respectively, and the total barred sample is shown as a black open histogram. All distributions are normalized.}
        \label{barpara}%
    \end{figure*}

The distributions of bar structural parameters for the barred disk galaxy sample are shown in Figure \ref{barpara}. The absolute bar length ($R_{\mathrm{bar}}$) shows a pronounced peak around 5~kpc, consistent with the typical physical scale of large-scale stellar bars observed in nearby disk galaxies \citep{2011MNRAS.415.3308G}. The normalized $R_{\mathrm{bar}}$ is mainly distributed between 0.2 and 0.7, indicating that most bars extend over roughly one-quarter to three-quarters of the optical disk.

The ellipticity distribution increases toward higher ellipticity, reaching a broad peak at $e_\mathrm{{bar}}\sim 0.4$, 
and then declines rapidly beyond $e_\mathrm{{bar}}>0.6$, with very few bars exhibiting $e_\mathrm{{bar}}>0.8$. The broad peak reflects a wide range of intrinsic bar strengths and orientations, rather than a sharply defined characteristic shape. 
At low ellipticities ($e_\mathrm{{bar}}<0.3$), the frequency remains moderate, likely due to projection effects and the inclusion of weak or nearly face-on bars with reduced apparent elongation. 
In contrast, the steep truncation at high ellipticity suggests that extremely elongated bars are intrinsically rare or obscured by dust extinction. Notably, strong bars have larger absolute and normalized $R_{\mathrm{bar}}$ than weak bars, and their $e_\mathrm{{bar}}$ are significantly higher. This difference is partly expected, because bars with high ellipticity are visually clearer and are therefore more likely to be classified as strong bars in the Galaxy Zoo scheme.

Overall, the distributions of bar length and ellipticity are consistent with those expected for disk-dominated galaxies at low redshift. In particular, \citet{2009A&A...495..491A} found that bar lengths derived from SDSS DR5 peak at $\sim 5$ kpc and decline towards $\sim 20$ kpc, while the distribution of normalized bar lengths reaches a maximum around $\sim 0.4$ and decreases towards $\sim 1.4$. These trends are in good agreement with the ranges and peak locations found in our sample, thereby supporting the reliability of our bar measurements.

        \begin{figure*}
        \centering
        \includegraphics[width=\textwidth]{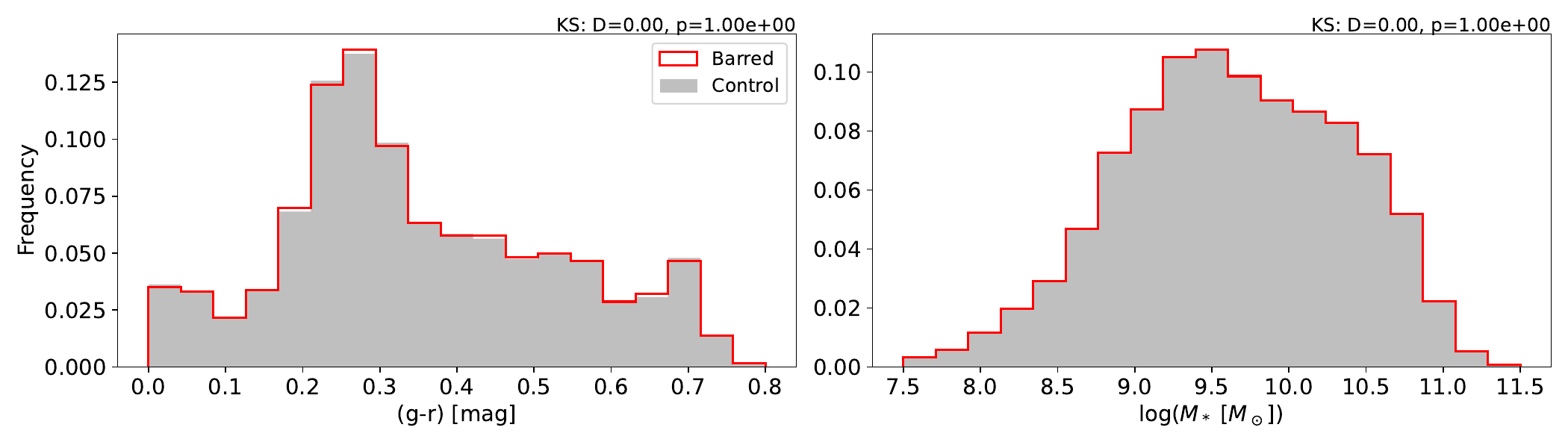}
        \caption{Distributions of $M_{\ast}$ and ($g-r$) for the barred (red open histogram) and control (gray filled histogram) samples. KS test results comparing the two distributions are indicated in the upper right corner of each panel.}
        \label{control}%
    \end{figure*}

       \begin{figure*}
        \centering
        \includegraphics[width=0.93\textwidth]{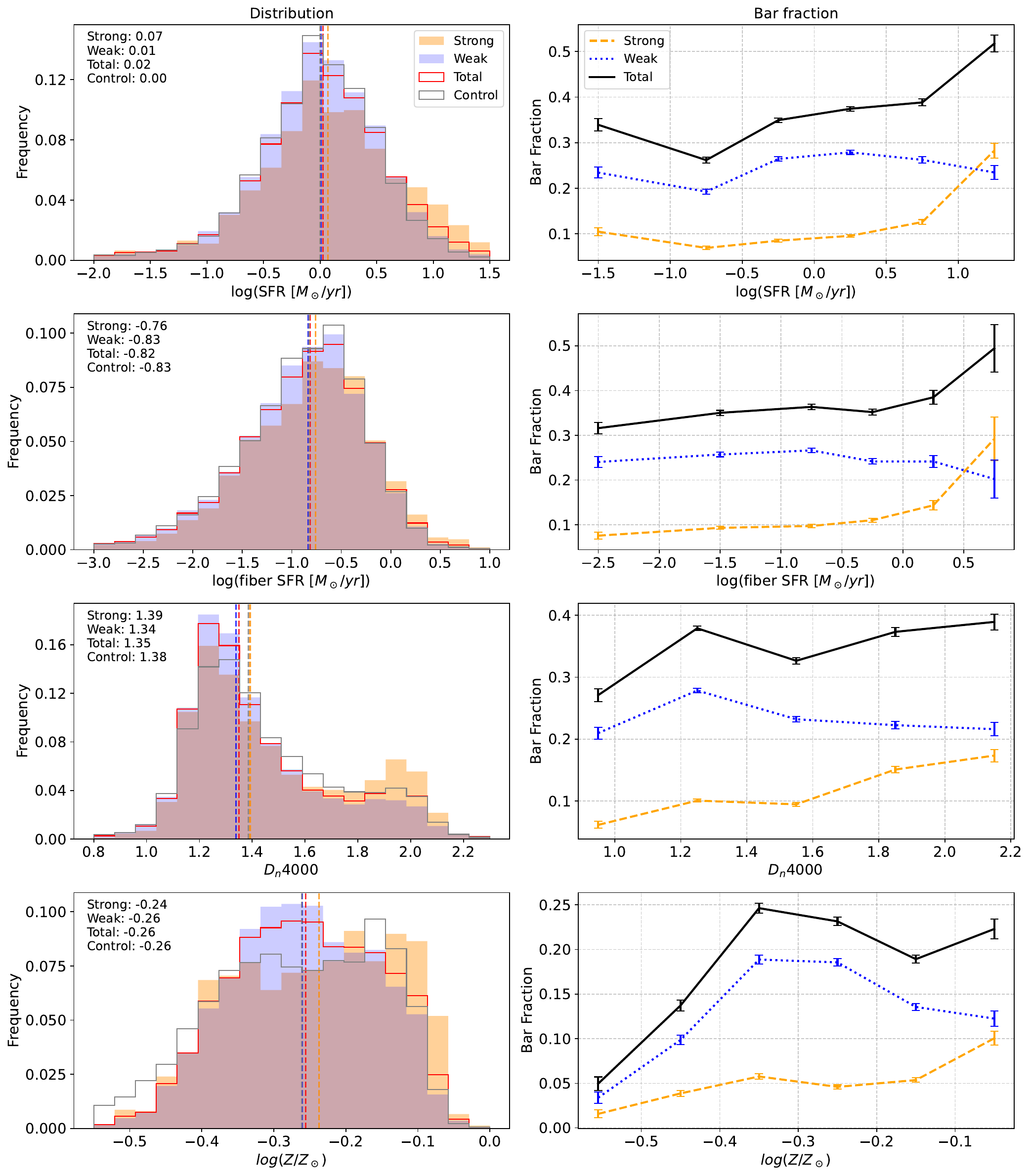}
        \caption{The distributions of physical parameters for disk galaxies in the barred and control samples. The left panels show the distributions of log(SFR), log(fiber SFR), $D_n(4000)$, and $Z/Z_\odot$ for strongly, weakly, total barred samples and control sample, shown as orange filled, blue filled, red open, and gray open histograms, respectively. The $Z/Z_\odot$ is computed only for star-forming galaxies. Dashed lines in the corresponding colors indicate the medians of the four subsamples. Median values are shown in the upper-left corner of each panel. The right panels show the dependence of the bar fraction on each physical parameter. Line styles are the same as in Figure~\ref{GR_M}. Error bars indicate the binomial uncertainties of the bar fraction in each bin.
}
        \label{other_para}%
    \end{figure*}

   \begin{figure*}
        \centering
        \includegraphics[width=0.93\textwidth]{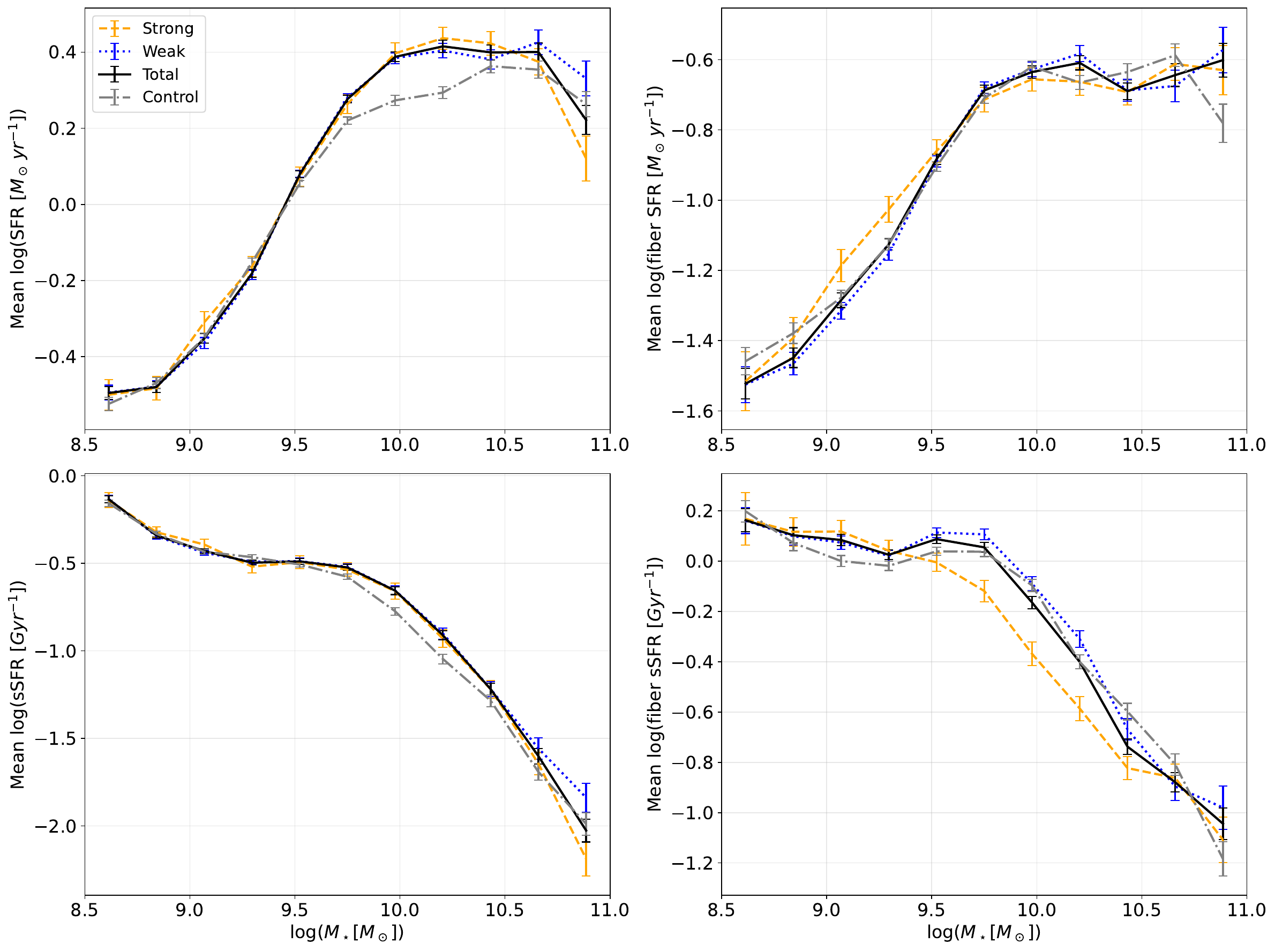}
        \caption{Mean global and fiber SFRs and sSFRs as a function of $M_{\ast}$, strongly, weakly, total barred samples and control sample are shown in orange dashed line, blue dotted line, black solid line, and gray dash-dotted line, respectively. Error bars indicate the standard errors.}
        \label{sSFR_M}%
    \end{figure*}

   \begin{figure*}
        \centering
        \includegraphics[width=0.93\textwidth]{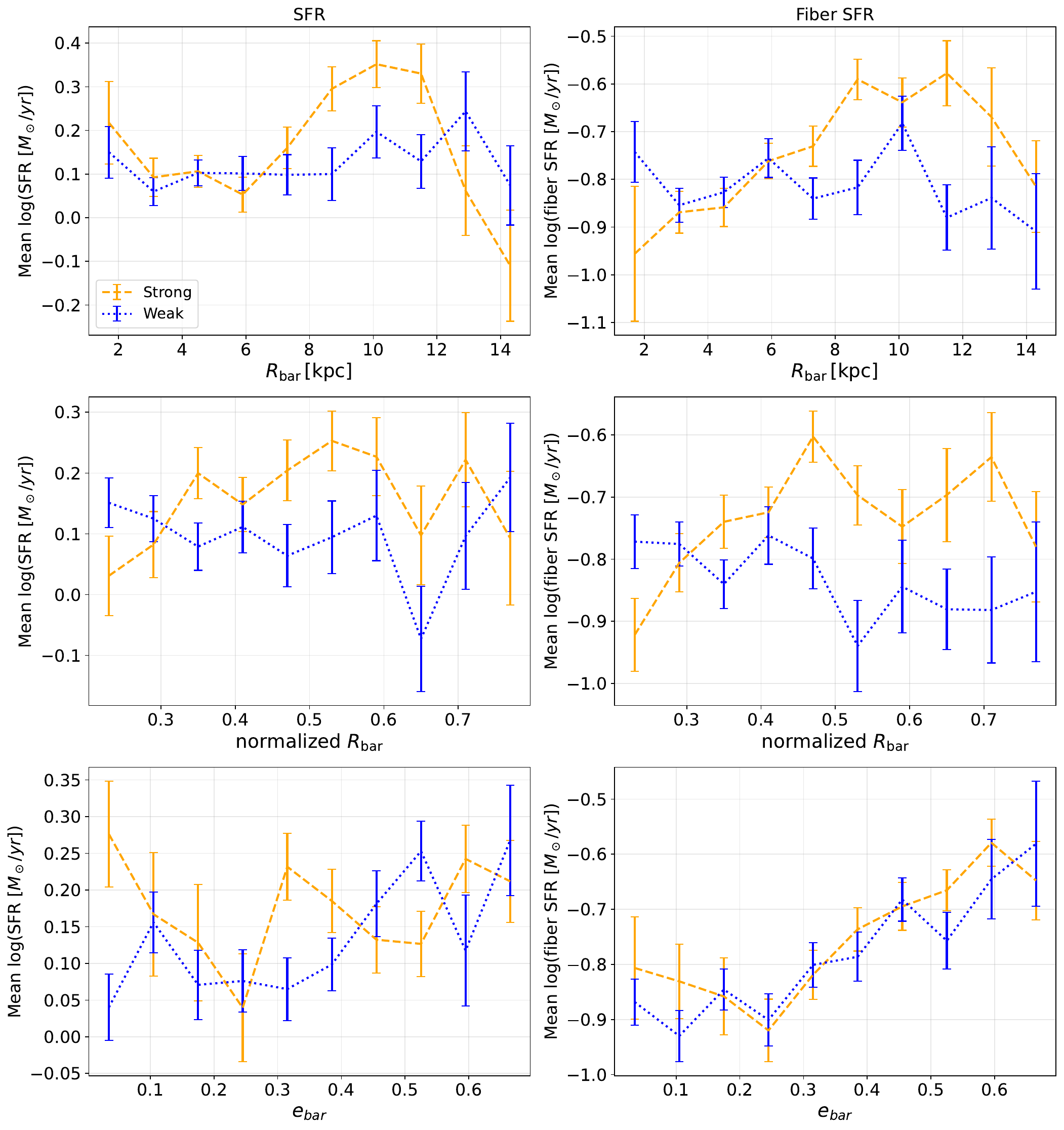}
        \caption{Mean global and fiber SFRs as a function of $R_{bar}$, normalized $R_{bar}$ and $e_{bar}$. Strongly and weakly barred samples are shown in orange dashed line and blue dotted line, respectively. The error bars indicate the standard errors of the mean SFR in each bin.}
        \label{meanSFR}%
    \end{figure*}

   \begin{figure*}
        \centering
        \includegraphics[width=0.93\textwidth]{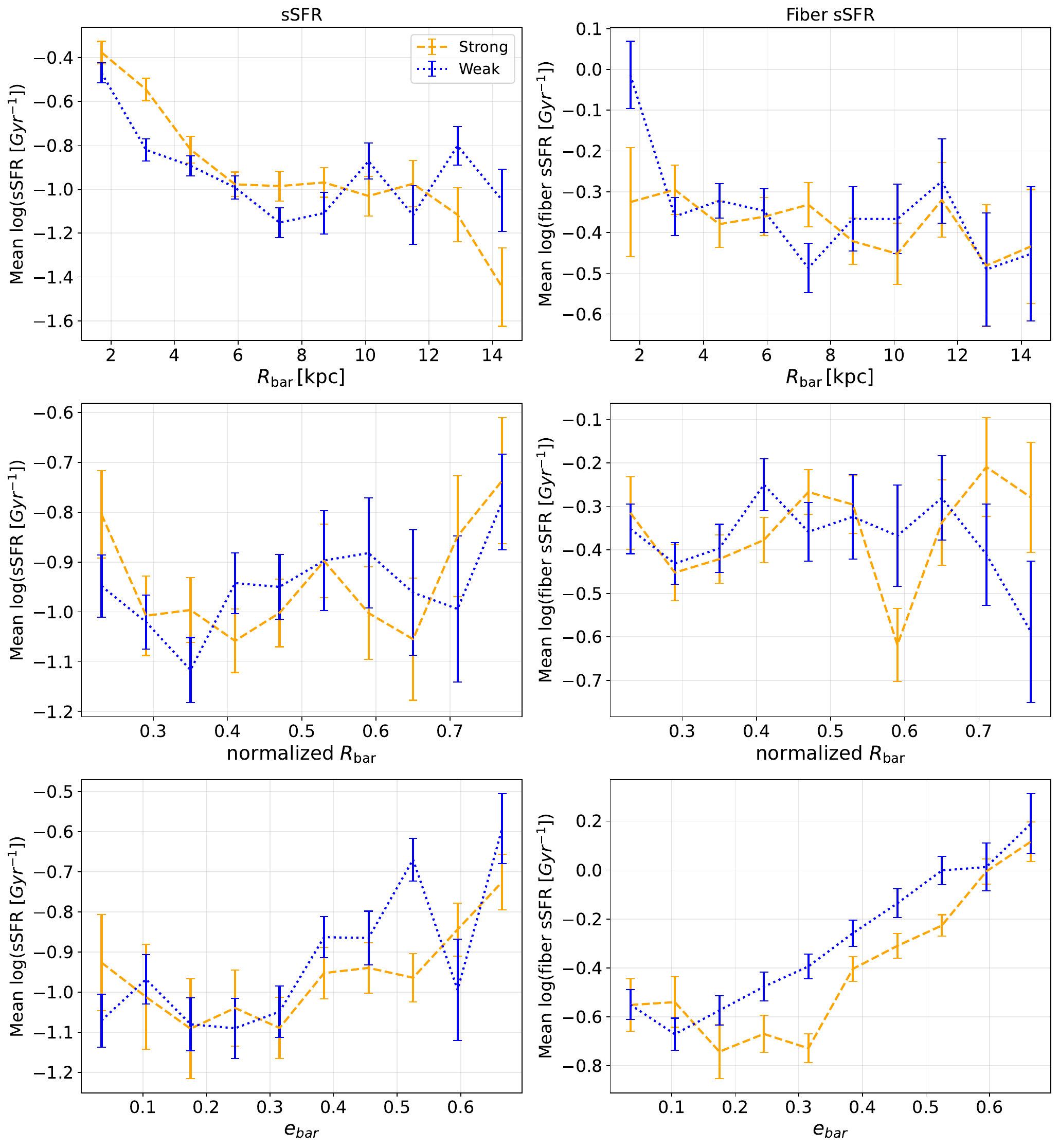}
        \caption{Relations between global and fiber sSFRs with bar structural parameters. Line styles are the same as in Figure~\ref{meanSFR}.}
        \label{sSFR_R}%
    \end{figure*}

\begin{figure*}
    \centering
    \includegraphics[width=0.93\textwidth]{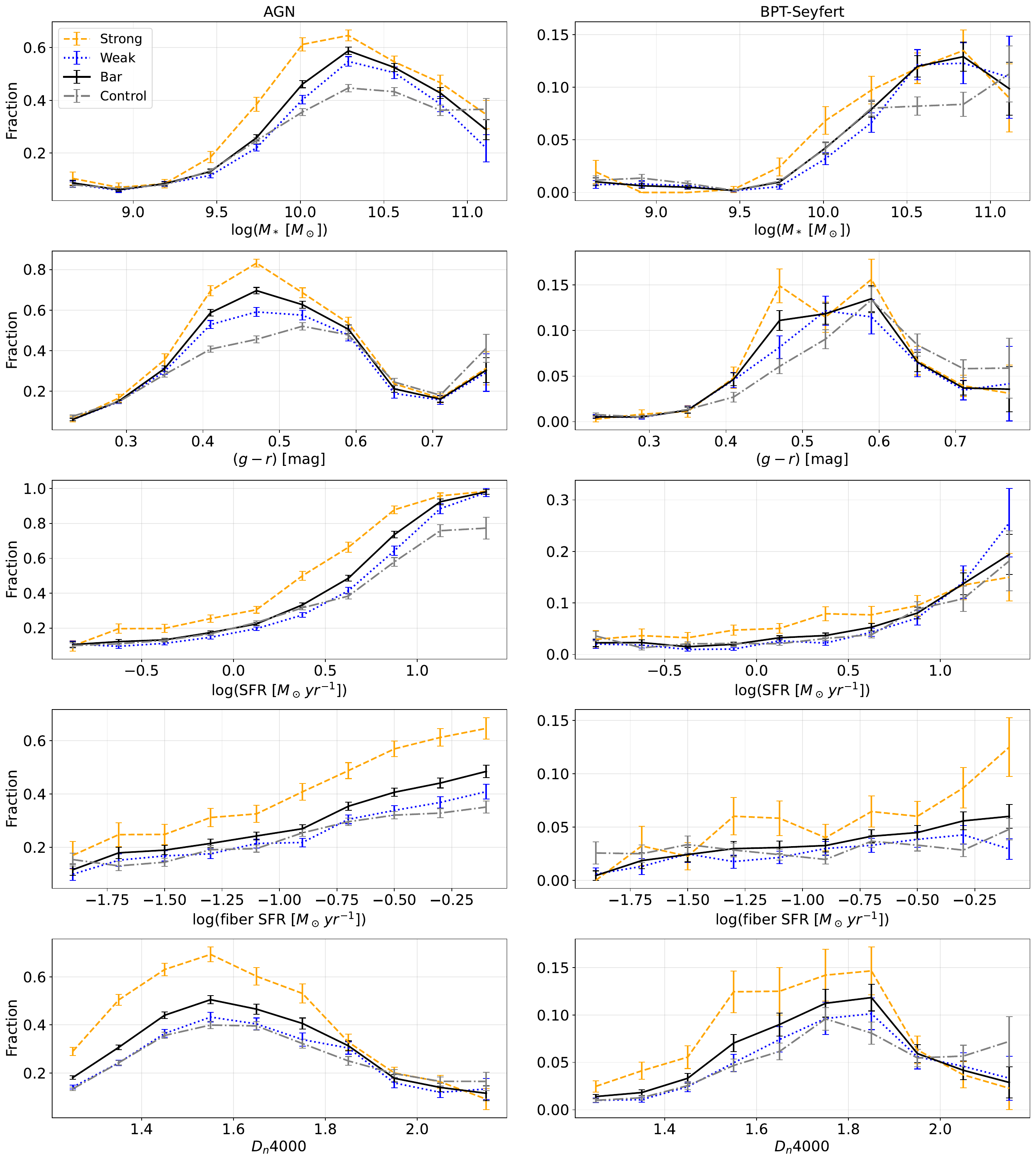}
    \caption{AGN fraction as a function of $M_\ast$, ($g-r$), SFR, fiber SFR, and $D_n4000$. The left panels show the fraction of galaxies identified as AGN by any diagnostic method, while the right panels show those classified as \textsc{Seyfert} by at least one BPT diagram. Strongly, weakly, total barred galaxies and galaxies in control sample are shown in orange dashed line, blue dotted line, black solid line, and gray dash-dotted line, respectively. Error bars indicate the binomial uncertainties of the fraction in each bin.}
    \label{bptagn_fraction_mult}
\end{figure*}

\begin{figure}
    \centering
    \includegraphics[width=0.45\textwidth]{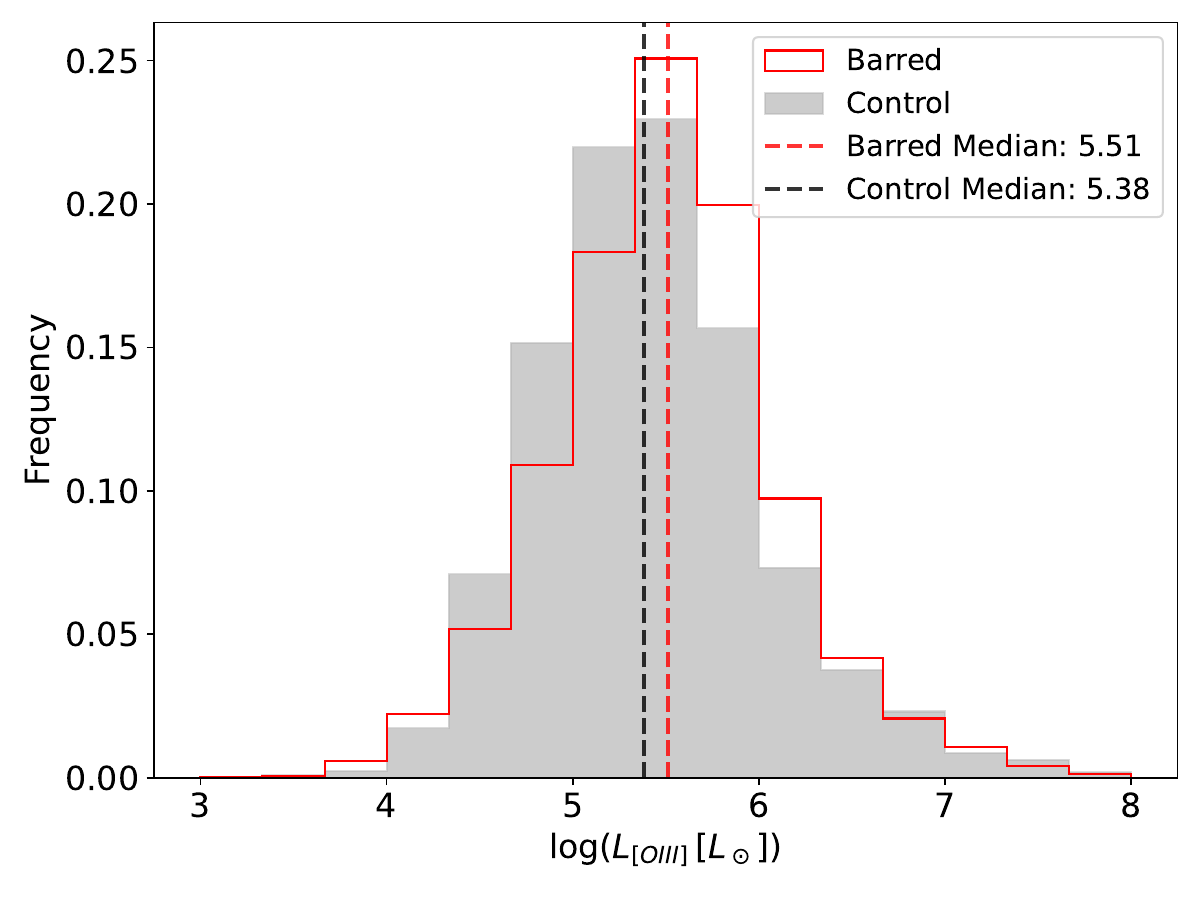}
    \caption{Distributions of [O III] $\lambda5007$ luminosity for AGN host galaxies in barred and control sample. The median $\log(L_{\text{[O III]}})$ of barred galaxies is shown in red dash line and unbarred galaxies in gray.}
    \label{oiii_luminosity}
\end{figure}

\begin{figure*}
    \centering
    \includegraphics[width=0.9\textwidth]{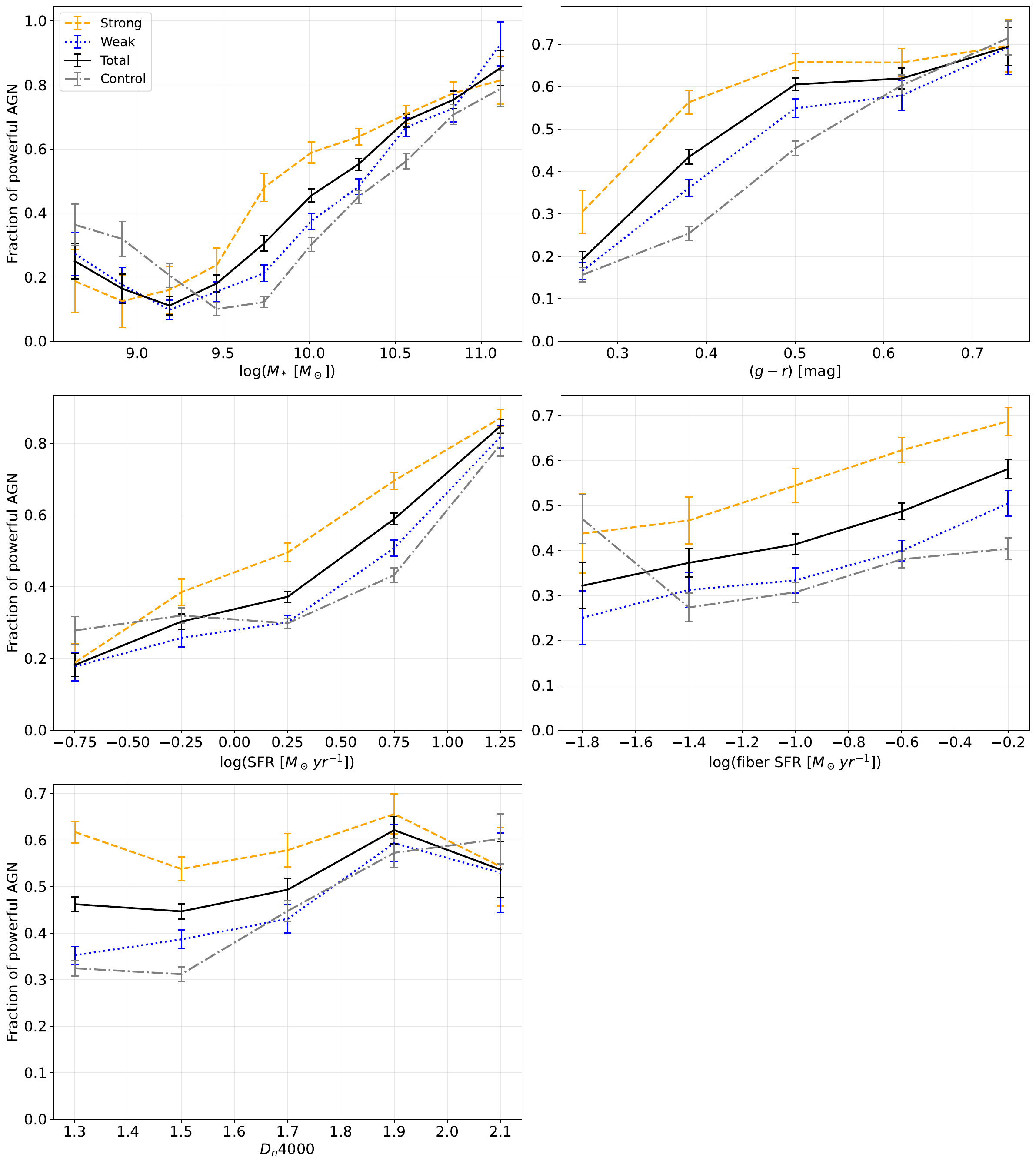}
    \caption{Fraction of AGN with $\log(L_{\text{[O III]}})>5.51\,L_\odot$ as a function of physical parameters: $\log M_*$, $(g-r)$, SFR, fiber SFR, and $D_n(4000)$. In each panel, the orange dashed lines, blue dotted lines, black solid lines, and gray dash-dotted lines represent strongly barred, weakly barred, total barred galaxies and galaxies in control sample, respectively. Error bars indicate the binomial uncertainties of the AGN fraction in each bin.}
    \label{agnfrac_multi}
\end{figure*}

\subsection{Connection with Star Formation Activity}
\label{subsubsec:sfr}

Differences in galaxy color and stellar mass may lead to statistical biases in SFR and other physical properties. To mitigate these effects, we constructed a control sample for the barred galaxies. All galaxies in the control sample are drawn from the unbarred sample, and are selected to closely match the barred galaxies in both color and stellar mass. Specifically, we exclude candidates whose $M_{\ast}$ differs by more than 0.1 deg or whose ($g-r$) differs by more than 0.05 mag from their barred counterparts. This matching constraint leads to the exclusion of some barred galaxies for which sufficiently similar unbarred counterparts are not available. The final control sample contains 11,444 galaxies. Figure~\ref{control} shows that the barred and control samples are effectively matched in stellar mass and color, as confirmed by KS tests (p = 1).
The left panels of Figure~\ref{other_para} show the distributions of global SFR, fiber SFR, $D_n(4000)$, and $Z/Z_{\odot}$ for the barred and control samples, with the corresponding KS statistics and $p$-values summarized in Table~\ref{tab:ks}, and the right panels show the bar fraction as a function of these parameters. We applied the KS test to compare the distributions of these physical properties between the barred and control samples.

As shown in Figure~\ref{other_para}, barred galaxies show systematically higher global SFRs than the control sample, although the enhancement is relatively modest. This offset is primarily driven by strongly barred systems, which tend to lie in galaxies with high SFR, whereas weakly barred galaxies show no significant difference relative to the control sample ,as confirmed by the KS test in Table~\ref{tab:ks}. Interestingly, when considering the central SFR, the difference between the total barred and control samples largely disappears. However, the strong bar sub-sample remains statistically distinguishable from its control, albeit with a reduced offset ($D=0.05$).
The strong and total bar fractions increase with global SFR at SFR $>10^{-0.5}M_{\odot}\,\mathrm{yr}^{-1}$, with the strong bar fraction rising from $<10\%$ to $\sim30\%$. A similar trend is observed for the fiber SFR, where the strong bar fraction increases from $<10\%$ at fiber SFR $\sim10^{-2.5}M_{\odot}\,\mathrm{yr}^{-1}$ to $\sim30\%$ at fiber SFR $\sim10^{0.7}M_{\odot}\,\mathrm{yr}^{-1}$. In contrast, the weak bar fraction remains nearly constant at $20\%$ -- $30\%$, suggesting that the weak bar fraction is largely independent of both global and fiber SFR.

In the relation between the mean total and fiber SFR and $M_{\ast}$, as shown in Figure~\ref{sSFR_M}, we find that at $M_{\ast} \sim 10^{9.5}-10^{10.5}\,M_{\odot}$, unbarred galaxies exhibit lower SFRs than barred galaxies, while no significant difference is found between strongly and weakly barred systems. Considering the fiber SFR, strongly barred galaxies with $M_{\ast} \sim 10^{9}-10^{9.5}\,M_\odot$ appear to show slightly elevated central SFRs compared to other subsamples, whereas for more massive systems, all subsamples display nearly identical fiber SFRs.

To mitigate potential biases arising from the strong dependence of SFR on stellar mass, we further investigate the specific star formation rate (sSFR). The global sSFR is computed using the total stellar mass $M_\ast$, whereas the fiber sSFR is derived using a fiber stellar mass estimated by scaling the total stellar mass with the ratio of the fiber $\it g$ band luminosity to the total galaxy $\it g$ band luminosity. The luminosities are corrected for galactic extinction based on the \citet{1998ApJ...500..525S} dust maps, with recalibration following \citet{2011ApJ...737..103S}. This procedure implicitly assumes a uniform mass-to-light ratio and may therefore introduce systematic uncertainties. Nevertheless, it provides a consistent way to compare central and global star formation efficiencies across different subsamples. 
As shown in Figure~\ref{sSFR_M}, the mean global sSFR remains broadly comparable among strongly barred, weakly barred, total barred, and unbarred control galaxies across nearly all stellar mass bins, except for a very slight decrease in the unbarred sample at $M_\ast \sim 10^{9.5}–10^{10.5}\,M_\odot$. However, subtle systematic deviations emerge when examining the fiber sSFR. While no significant difference in fiber sSFR is observed between weakly barred and unbarred galaxies, at $M_\ast \sim 10^{9.5}-10^{10.5}\,M_\odot$, strongly barred galaxies exhibit a deficit of $\sim$ 0.2 dex in fiber sSFR relative to other subsamples. This mass-dependent behavior may reflect the role of bars in galaxy evolution. In massive systems, strong bars can efficiently redistribute angular momentum and drive gas toward the central regions over extended timescales, this process not only exhausts the available gas reservoir but also leads to morphological quenching, with the growth of the central bulge, resulting in the observed lower sSFR.

We further examined how the mean global and fiber SFRs correlate with the structural parameters of bars. To mitigate the potential impact of stellar-mass differences between strong and weak bars, we constructed a stellar mass–matched control sample for the strongly barred galaxies drawn from the weakly barred population, following the same procedure described above. Since some strongly barred galaxies have no suitable counterparts in the weakly barred population, the final weakly barred control sample contains 2418 galaxies.
As shown in Figure~\ref{meanSFR}, weakly barred galaxies show no clear dependence of either global or fiber SFR on absolute or normalized bar length. For the strongly barred sample, the global and fiber SFRs exhibit no significant dependence on normalized bar length, but exhibit a non-monotonic dependence on the absolute bar length. In particular, both global and fiber SFR are elevated in the range of $\sim8-12$ kpc compared to other intervals. Over the range of $\sim2-8$ kpc, the fiber SFR shows a positive correlation with increasing bar length. However, at bar lengths $>12$ kpc, both global and central SFR decline rapidly.

Turning to the ellipticity, no significant correlation is found between global SFR and bar $e_{bar}$ for either strong or weak bars. However, fiber SFR increases with bar ellipticity within the typical range of $e_{bar} \sim 0.2-0.6$ \citep[][]{2025ApJ...982..129W}, and at fixed ellipticity, strongly and weakly barred disks exhibit nearly identical mean fiber SFR, although strongly barred systems showing a marginally higher level, suggesting that the central SFR is more closely tied to bar ellipticity than to the strong, weak bar classification, and bar ellipticity is likely more intrinsically coupled to the bar's non-axisymmetric potential.

When we turn to the sSFR, a different trend emerges. As shown in Figure~\ref{sSFR_R}, the global sSFR exhibits an overall anti-correlation with $R_{\rm{bar}}$, whereas the fiber sSFR shows no dependence on $R_{\rm{bar}}$. No clear correlation is found between sSFR and normalized $R_{\rm{bar}}$. In contrast, fiber sSFR displays a strong positive correlation with bar ellipticity, similar to the behavior seen for fiber SFR in Figure~\ref{meanSFR}. At fixed ellipticity, strongly and weakly barred galaxies exhibit very similar fiber sSFR, with weakly barred systems showing a marginally higher level. This divergence between SFR and sSFR trends confirms that while strong, elongated bars are efficient at channeling gas, physically larger bars are often hosted by more evolved, lower-sSFR galaxies.

\subsection{Central Stellar Population Age and Metallicity}
\label{Dn4000 and Z/Z}

As shown in the third row of Figure~\ref{other_para}, all subsamples display a pronounced peak near $D_n4000 = 1.3$, corresponding to relatively young stellar populations. The distribution of weakly barred galaxies is particularly concentrated at this peak, resulting in a higher frequency of weakly barred systems compared to the control sample. Strongly barred galaxies show an additional, smaller peak around $D_n4000 = 2.0$, producing a bimodal distribution in both strong and total bar samples. The fraction of strongly barred galaxies increases steadily with $D_n4000$, whereas the weak bar fraction rises below $D_n4000 \sim 1.3$ but declines at higher values. Overall, the increasing trend of the total bar fraction indicates that barred galaxies, especially strong bars, tend to host older stellar populations.

The metallicity analysis is restricted to galaxies classified as star-forming. After applying the classification according to the [N II] based BPT diagnostic in the AGN/Galaxy Classification VAC, we retain only galaxies classified as star-forming, excluding composite and AGN (including Seyfert and LINER) systems, ensuring that the metallicity measurements are based on a clean star-forming sample and are not biased by potential AGN contamination effects. The final barred star-forming sample contains 5,979 galaxies. A corresponding control sample is then reconstructed from unbarred star-forming galaxies following the same stellar mass and color matching procedure described in Section~\ref{subsubsec:sfr}, resulting in 5,730 matched counterparts.
As shown in the bottom row of Figure~\ref{other_para}, strongly barred star-forming galaxies exhibit the highest median metallicity ($\log(Z/Z_\odot)=-0.24$), while the other subsamples have very similar median values clustered around $\log(Z/Z_\odot)\sim-0.26$, although the KS test indicates that the weak bar and control samples are statistically distinguishable despite the small difference (see Table~\ref{tab:ks}). In addition, strongly barred systems show an enhanced frequency at the metal-rich end of the distribution ($\log(Z/Z_\odot)>-0.15$) compared to the other subsamples, whereas the control sample shows the highest frequency at the low metallicity end ($\log(Z/Z_\odot)<-0.4$). The strong bar fraction increases with metallicity, while the fraction of weak bars exhibits a noticeable decline at high metallicity ($\log(Z/Z_{\odot})>-0.35$). These trends underscore a strong correlation between bar prominence and the degree of chemical enrichment in the host galaxy, suggesting that strong bars are preferentially hosted by chemically evolved systems.

\subsection{Connection with AGN Activity}

To explore how the presence of bars relates to nuclear activity, we examine the connection between bars and AGN fractions in disk galaxies. The DESI DR1 AGN/Galaxy Classification VAC includes multiple methods to identify AGNs, including BPT diagnostics, infrared WISE color diagnostics, and other optical and UV line diagnostics. These AGN identification methods are organized into four categories: (1) BPT-\textsc{Seyfert}, specifically classified as \textsc{Seyfert} galaxies by at least one of the three standard BPT diagnostic diagrams based on [N~II], [S~II], or [O~I]; (2) BPT-ANY AGNs, identified as \textsc{Seyfert}, LINER, or Composite by at least one BPT diagnostic; (3) WISE AGNs, selected via at least one infrared (WISE) color diagnostic, the WISE diagnostics cuts are adopted from \cite{2011ApJ...735..112J}, \cite{2012MNRAS.426.3271M}, \cite{2012ApJ...753...30S}, \cite{2018ApJS..234...23A}, \cite{2020ApJ...903...91Y} and \cite{2022AJ....163..224H}; (4) Other AGNs, identified by other optical diagnostics, include the [He~II] BPT diagram \citep[][]{2012MNRAS.421.1043S}, the WHAN diagram \citep{2011MNRAS.413.1687C}, the BLUE diagram \citep{2004MNRAS.350..396L,2010A&A...509A..53L}, the Mass-Excitation diagram \citep{2014ApJ...788...88J}, and the Kinematics-Excitation diagram \citep{2018ApJ...856..171Z}. The fractions of overall AGNs and each AGN subcategory are reported in Table~\ref{table:AGN_frac}. For clarity and to facilitate comparison with previous work, we focus our subsequent analysis on the overall AGN fraction and the BPT AGN fraction. As shown in Table~\ref{table:AGN_frac}, the barred galaxies, especially those with strong bars, consistently show higher AGN and BPT-\textsc{Seyfert} fractions than the control sample. Strong bar sample shows higher AGN as well as BPT-\textsc{Seyfert} fractions than weak bar sample, while the weakly barred galaxies exhibit slightly lower overall AGN and BPT-\textsc{Seyfert} fractions compared to the unbarred systems.

\begin{table}[h]
\centering
\caption{AGN fractions and counts for different galaxy subsamples.}
\label{table:AGN_frac}
\setlength{\tabcolsep}{3pt}
\begin{tabular}{lcccccc}
\hline\hline
Sample & $N_{\mathrm{total}}$ & Overall AGN & BPT-\textsc{Seyfert} & BPT-Any & WISE AGN & Other AGN \\
\hline
Disk total      & 33,201 & 23.8\% (7,893) & 3.3\% (1,081) & 11.5\% (3,833) & 0.9\% (300) & 22.8\% (7,587) \\
Barred          & 11,444 & 27.1\% (3,099) & 3.7\% (424)   & 14.5\% (1,659) & 0.8\% (88)  & 26.5\% (3,033) \\
Strongly barred & 3,387  & 37.9\% (1,285) & 5.9\% (200)   & 23.7\% (802)   & 1.0\% (33)  & 37.3\% (1,265) \\
Weakly barred   & 8,057  & 22.5\% (1,814) & 2.8\% (224)   & 10.6\% (857)   & 0.7\% (55)  & 21.9\% (1,768) \\
Control         & 11,444 & 23.3\% (2,671) & 3.2\% (370)   & 11.8\% (1,349) & 0.6\% (74)  & 22.7\% (2,594) \\
\hline
\end{tabular}
\end{table}

Figure~\ref{bptagn_fraction_mult} presents the AGN fractions for each subsample as a function of several physical properties. As shown in Figure~\ref{bptagn_fraction_mult}, barred galaxies exhibit higher AGN and BPT-\textsc{Seyfert} fractions than unbarred galaxies at stellar masses above $M_{\ast}\sim 10^{10} M_{\odot}$. Strongly barred galaxies show elevated AGN fractions already at lower masses, and display a pronounced excess in the intermediate mass range $M_{\ast}\sim10^{9.5}$–$10^{10.5}M_{\odot}$, nearly twice that of unbarred galaxies at $M_\ast \sim 10^{10}\,M_\odot$. In terms of color, barred systems show higher AGN and BPT-\textsc{Seyfert} incidence in the range $0.35 < g-r < 0.6$.
Conversely, the disparity  diminishes at the red end, and unbarred galaxies displaying comparable overall AGN fractions and even higher BPT-\textsc{Seyfert} fractions than their barred counterparts.
This suggests that the enhancement of AGN activity driven by bars is most pronounced during the transition through the green valley, supporting the role of bars in accelerating both nuclear accretion and host galaxy quenching.

Regarding star formation, the AGN fraction exhibits a monotonic increase with both global and fiber SFRs across all subsamples. At low global SFRs ($\sim 10^{0.5}\,M_\odot\ \mathrm{yr}^{-1}$), the overall AGN fractions of barred and unbarred galaxies are similar, although strongly barred galaxies tend to have higher AGN fractions than weakly barred or unbarred galaxies at a given SFR. At high SFR, both strong and weak bars are associated with enhanced AGN and BPT-\textsc{Seyfert} fractions relative to unbarred galaxies. For fiber SFR, strong bar sample exhibit a higher fraction of both AGN and BPT-\textsc{Seyfert} than unbarred sample across almost the entire range, while weak bar galaxies are nearly indistinguishable from those of the unbarred galaxies.
This suggests that strong bars likely provide a more sufficiently non-axisymmetric potential to consistently channel gas into the nuclear region.

The relation with stellar age shows that barred galaxies, especially those with strong bars, have higher AGN and BPT-\textsc{Seyfert} fractions at $D_n4000 < 1.9$. For the oldest systems ($D_n4000 > 1.9$) the overall AGN incidence falls markedly and the differences between morphological classes disappear. Collectively, these results suggest that while bars are efficient fuel-delivery mechanisms, their ability to trigger AGN is fundamentally modulated by the host galaxy’s gas content and evolutionary state.

We now focus our analysis on the incidence of more powerful AGNs, selected based on [O III]~$\lambda5007$ luminosity ($L_{\text{[O III]}}$). The [O III]~$\lambda5007$ emission line is one of the most widely used optical diagnostics of AGN activity. It originates from the narrow-line region (NLR), which is photoionized by the hard radiation field of the central accreting black hole \citep{BPT1981PASP...93....5B, 2006MNRAS.372..961K(OIII), 2004ApJ...613..109H(OIII)}. Since the NLR is spatially extended and less affected by dust obscuration than the broad-line region, $L_{\text{[O III]}}$ provides a robust tracer of the ionizing power of the AGN \citep{2003AJ....126.2125Z, 2015MNRAS.454.3622B, 2017MNRAS.467..540L}. Higher $L_{\text{[O III]}}$ typically correspond to more powerful AGN, though moderate contributions from star formation or extinction may introduce some scatter.

Figure~\ref{oiii_luminosity} shows the distributions of $L_{\text{[O III]}}$ for AGN host galaxies in barred and control samples. Barred galaxies exhibit slightly higher $L_{\text{[O III]}}$ values compared to unbarred ones, showing a higher frequency toward the luminous end of the distribution. The median log($L_{\text{[O III]}}$) of barred galaxies is $5.51\,L_\odot$, based on this, we classify galaxies with $\log(L_{\text{[O III]}})>5.51\,L_\odot$ as hosting powerful AGN. A similar criterion was adopted by \citet{2025A&A...699A.204M}, who defined powerful AGN using the median $\log(L_{\text{[O III]}})>6.2\,L_\odot$ of their barred disk sample. By differentiating AGN based on their [O III] luminosity, we can disentangle whether bars merely trigger low-level nuclear activity or are capable of fueling the most luminous and energetically significant accretion events.

Figure~\ref{agnfrac_multi} shows the fraction of powerful AGN ($\log(L_{\text{[O III]}})>5.51\,L_\odot$) as a function of stellar mass $M_\ast$, color $(g - r)$, total and fiber SFR, and $D_n4000$. The observed trends largely mirror those of the overall AGN population discussed in Figure~\ref{bptagn_fraction_mult}. Across all relations, strongly and weakly barred galaxies exhibit broadly similar trends, yet the fraction of powerful AGN is generally higher in strongly barred systems. 
For galaxies with $M_\ast < 10^{9.5}\,M_\odot$, unbarred systems exhibit a comparable or marginally higher fraction of powerful AGN than barred galaxies. In contrast, for $M_\ast > 10^{9.5}\,M_\odot$, barred galaxies consistently show a higher fraction of powerful AGN than their unbarred counterparts, and this fraction increases with stellar mass. In terms of color, the fractions of powerful AGN increase toward redder galaxies, barred galaxies exhibit a higher powerful AGN fraction in the range $0.2 < (g - r) < 0.6$, while the difference in the fraction between barred and unbarred AGN hosts diminishes for $(g-r) > 0.6$. For global SFR, powerful AGN are preferentially hosted by galaxies with higher SFRs, unbarred galaxies exhibit a comparable or slightly higher powerful AGN fraction than barred galaxies at $\mathrm{SFR} < 1\,M_\odot\ \mathrm{yr}^{-1}$, whereas barred galaxies become more dominant at higher SFR. The fiber SFR shows a similar trend: unbarred galaxies have a slightly higher AGN fraction at low fiber SFR, but the relation reverses for $\mathrm{fiber\,SFR} > 10^{-1.6}\,M_\odot\ \mathrm{yr}^{-1}$. This reversal likely reflects that at low SFR, AGN activity in unbarred systems can be sustained by secular or external processes, while at high SFR, the presence of bars facilitates efficient gas inflow, simultaneously enhancing both star formation and AGN fueling. 

In terms of stellar age, galaxies with $D_n4000 < 1.8$ show a clear excess of powerful AGN among barred systems relative to unbarred AGN hosts. Within this regime, strongly barred galaxies exhibit a markedly higher fraction of powerful AGN than unbarred galaxies, whereas weakly barred systems display only a modest enhancement. The contrast in the observed fractions becomes more pronounced toward younger stellar populations. In contrast, at $D_n4000 > 1.8$, the presence of bars shows no obvious association with the incidence of powerful AGN.

\begin{figure*}
    \centering
    \includegraphics[width=0.9\textwidth]{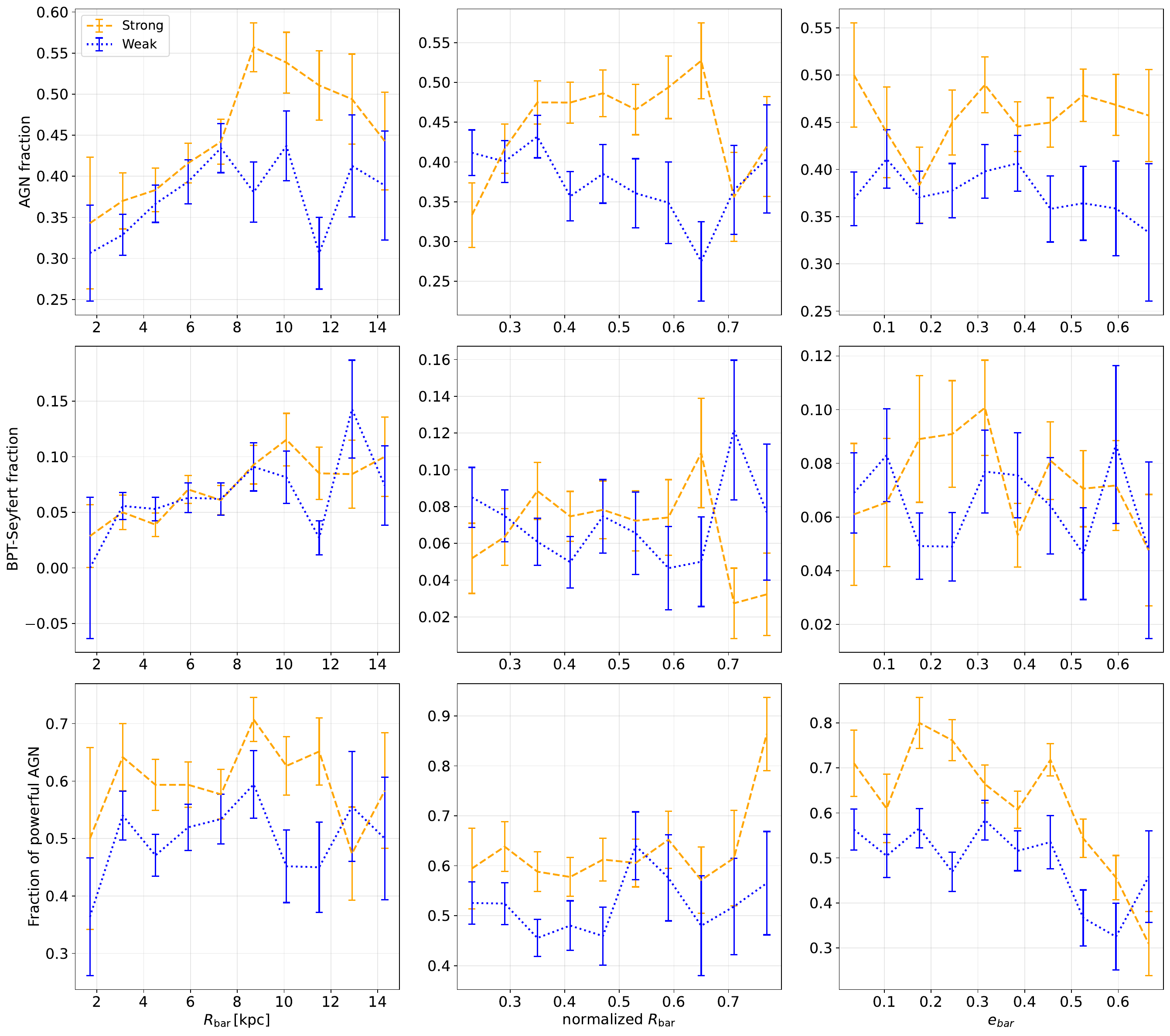}
    \caption{The top, middle, and bottom panels show the fractions of AGN, BPT-\textsc{Seyfert}, and powerful AGN, respectively, as functions of $R_{\mathrm{bar}}$, normalized $R_{\mathrm{bar}}$, and $e_{\mathrm{bar}}$. In each panel, the orange dashed lines and blue dotted lines represent strongly and weakly barred galaxies, respectively. The error bars indicate the binomial uncertainties of the AGN fraction in each bin.}
    \label{agnfrac_barpara}
\end{figure*}

We further investigate the connection between AGN activity and bar structural parameters, as shown in Figure~\ref{agnfrac_barpara}. Overall, strongly barred galaxies exhibit significantly higher fractions of AGN and BPT-\textsc{Seyfert} compared to weakly barred systems, and the fraction of powerful AGN is also clearly elevated among AGN hosts with strong bars relative to those with weak bars.
For strongly barred galaxies, the AGN fraction increases with bar length at $R_\mathrm{bar} < 8$ kpc and then declines at larger bar lengths. In contrast, weakly barred galaxies show a similar rising trend only over the range $R_\mathrm{bar} < 7$ kpc, beyond which the AGN fraction exhibits little dependence on bar length. The BPT-\textsc{Seyfert} fraction shows generally weak correlations with bar length, we observe only a mild increase for strongly barred systems with $R_\mathrm{bar} < 10$ kpc, from $\sim5\%$ to $\sim10\%$. We find no clear evidence that the fraction of powerful AGN depends on bar length. In particular, the powerful AGN fraction in strongly barred hosts remains relatively stable at $\sim60-70\%$ over the range $R_\mathrm{bar} \sim 4–12$ kpc, while weakly barred galaxies maintain a level of around $\sim50\%$. The dependence of AGN fraction and powerful AGN fraction on normalized bar length is also weak. Regarding bar ellipticity, there is little correlation with AGN or BPT-\textsc{Seyfert} fraction. Among strongly barred AGN hosts, the fraction of powerful AGN remains above 60\%, but decreases steadily in systems with high ellipticity ($e_\mathrm{bar} > 0.5$).

Taken together, these results show that strongly barred galaxies are, on average, more likely to host AGN and BPT-selected \textsc{Seyfert} than weakly barred and unbarred systems. Moreover, barred AGN hosts particularly those with strong bars, tend to show higher incidences of powerful AGN, especially within the most densely populated regions of parameter space. One possible interpretation is that strong bars maybe more efficiently redistribute angular momentum and drive gas toward central regions, increasing the availability of fuel for nuclear accretion and star formation \citep[][]{1989Natur.338...45S, 2000ApJ...529...93K}. Notably, although bars are clearly associated with both the AGN incidence and the fraction of powerful AGN, the correlations between AGN and bar structural parameters are relatively weak.

\section{Discussion}
\label{sec:discussion}

Our analysis of the DESI sample reveals that bars, especially strong bars, are preferentially hosted by massive, red, and evolved disk galaxies. Strongly barred galaxies exhibit slightly enhanced global and fiber SFR, yet the presence of a bar seems to have little effect on the sSFR. Among the structural parameters of bars, both SFR and sSFR show positive correlations with $e_\mathrm{bar}$, while galaxies with strong bars systematically display lower sSFR than those with weak bars. We further investigate the connection between bars and AGN and find that barred galaxies present a systematically higher AGN incidence. Among AGN hosts, the fraction of strong AGN is also higher in barred systems. However, the AGN activity appears only weakly related to either bar length or bar ellipticity. Taken together, these findings indicate that bars play a significant role in regulating the secular evolution of disk galaxies.

\subsection{Bar Effect on Star Formation}

Our results show that barred galaxies present a clear bimodality in the color–mass plane. Strong bars are common in intermediate to high mass systems ($M_\ast \sim 10^{10.5}\,M_\odot$) with red colors ($g-r \sim 0.6$), while weak bars are preferentially found in lower mass galaxies ($M_\ast \sim 10^{9.5}\,M_\odot$) with bluer colors ($g-r < 0.3$). Consistently, the bar fraction also increases toward higher metallicity, and strongly barred galaxies show a pronounced peak around $D_n4000 \sim 2$. These results align well with previous work. \citet{2016A&A...595A..63V} showed that strongly barred spirals tend to be redder, more metal-rich, and older than weakly barred or unbarred disks, consistent with accelerated evolution in barred systems, while \citet{2025MNRAS.542..151M} reported a bimodal distribution of barred galaxies in stellar mass and color. These trends indicate that galaxies hosting bars, especially strong bars, are on average more evolved than unbarred systems. This behavior may imply that either bars likely grow as galaxies evolve\citep[][]{2022MNRAS.512.5339R}, or more evolved disks are more likely to develop strong bars\citep[][]{2012ApJ...757...60K}.

Our analysis of star formation provides important insights into this evolutionary sequence. As shown in Figure~\ref{other_para}, the global SFR and sSFR distributions of barred and unbarred galaxies are broadly similar across most of the stellar mass range. However, within $M_* \sim 10^{9.5}-10^{10.5}\,M_{\odot}$, as shown in Figure~\ref{sSFR_M}, barred galaxies show a marginally higher global SFR but only a minimal increase in sSFR, suggesting that bars do not strongly affect the overall star formation budget. For central regions, we find that strongly barred galaxies show a enhancement in fiber SFR at $M_* \sim 10^{9}-10^{9.5}\,M_{\odot}$, which provides tentative indications that bars may contribute to redistributing star formation activity. In particular, bar-driven gas inflows may enhance star formation in the central regions, although this effect appears to be modest, not uniformly observed across all mass ranges, and dependent on galaxy properties such as stellar mass. A possible interpretation is that the bar may drive gas inflow, which transports cold gas into the inner disk and supplies fuel for central star formation\citep[][]{2019MNRAS.484.5192C}. \citet{2021MNRAS.507.4389G} found that star-forming galaxies hosting strong bars exhibit significantly higher fiber SFR, whereas this enhancement is notably absent in their quiescent counterparts. Studies of central SF “turnovers”, defined as central departures from the otherwise linear radial trends of stellar population diagnostics, characterized by a decrease in $D_n(4000)$ together with enhancements in EW(H$\delta_A$) and EW(H$\alpha$), indicate that $\sim$90\% of such galaxies host bars with order-of-magnitude enhancements in central SFR \citep[][]{2020MNRAS.499.1406L}.
By contrast, in more massive systems, within $M_* \sim 10^{9.5}-10^{10.5}\,M_{\odot}$ as shown in Figure~\ref{sSFR_M}, strongly barred galaxies exhibit a lower central sSFR than the other subsamples. Similarly, \citet{2023ApJ...943....7M} found that gas-poor galaxies with long bars exhibit lower sSFR than the star-forming main sequence at $M \sim 10^{10.5}\,M_{\odot}$. This trend can be understood if strong bars efficiently drive gas toward the central regions and trigger earlier episodes of star formation, thereby substantially increasing the stellar mass enclosed within the fiber aperture, particularly in massive galaxies at late evolutionary stages. Consequently, even if the star formation efficiency remains high, the enhanced stellar mass leads to a reduced central sSFR compared to other systems. This result likely reflects the accumulated influence of bars over long evolutionary timescales, rather than an immediate enhancement of star formation driven by the bar.

To better understand the role of bar structure, we further examine how bar-driven star formation depends on bar structural parameters. We find that fiber SFR and sSFR correlate more strongly with bar ellipticity $e_\mathrm{bar}$ rather than bar length, similar to the found in \citet{2012MNRAS.423.3486W}. This behavior is consistent with the theoretical expectation that more elongated bars induce stronger non-axisymmetric gravitational torques and thus drive gas more efficiently toward the center \citep[][]{2022A&A...666A.175Y}. However, the influence of bar length reveals a more nuanced evolutionary picture. For strongly barred galaxies, we find that both global and fiber SFR exhibit a non-monotonic dependence on absolute bar length. In the range of $R_\mathrm{bar} \sim 2-8$ kpc, the fiber SFR increases with bar length, which likely reflects an active growth phase where expanding bars exert progressively more efficient gas inflow \citep[][]{2003MNRAS.341.1179A}. The star formation activity reaches a peak at $R_\mathrm{bar} \sim 8-12$ kpc, whereas at $R_\mathrm{bar} >12$ kpc, both global and central SFR decline rapidly. This decline potentially marks a more evolved stage where gas reservoirs have been significantly depleted or redistributed \citep[][]{2005ApJ...632..217S}. Notably, at fixed bar ellipticity, weakly barred galaxies generally have higher sSFR, suggesting that strongly barred systems may be transitioning toward quenching. One possible scenario is that strong bars trigger more concentrated central star formation, accelerating gas consumption and chemical enrichment, which eventually leads to the depletion of the gas reservoir and the formation of central bulges \citep[][]{2017MNRAS.465.3729S}. In this sense, bars may play a dual role in galaxy evolution: they can enhance star formation on short timescales, and may also promote quenching on longer timescales \citep[][]{2021MNRAS.507.4389G}. \citet{2020MNRAS.499.1116F} showed that HI gas fractions are significantly lower in high-mass barred galaxies, and \citet{2021MNRAS.507.4389G} argued that enhanced fiber SFR accelerates quenching, these conclusions are consistent with our results on the connection between bars and star formation activity.

Besides, we note that fiber SFR shows almost no dependence on the strong/weak bar classification. This result highlights the limitations of binary morphological classes. It emphasizes the importance of continuous structural parameters such as bar ellipticity, which provide a more physically meaningful framework for interpreting the role of bars in regulating star formation \citep[][]{2002MNRAS.337.1118L}.

\subsection{Bar effect on AGN activity}

Our results related to AGN show that barred disks host a higher AGN fraction than unbarred systems, with this enhancement being primarily driven by the strongly barred subsample. The fraction of \textsc{Seyfert} selected from the BPT diagram is also systematically higher in strongly barred galaxies. When considering AGN strength, we find that AGN hosts with bars exhibit systematically higher $L_{\text{[O III]}}$ than the unbarred sample. In addition, the presence of a bar increases the fraction of powerful AGN over most parameter ranges, with the enhancement being most significant for strongly barred galaxies.
These findings are consistent with previous extragalactic studies. \citet{2025A&A...699A.204M} similarly report higher powerful AGN fractions in barred galaxies, and our finding that strongly barred galaxies exhibit a higher BPT-\textsc{Seyfert} fraction is also consistent with the analysis of \citet{2024MNRAS.532.2320G}.
We note that our BPT-\textsc{Seyfert} fractions show a noticeable discrepancy compared to those reported by \citet{2024MNRAS.532.2320G}. However, when restricting our sample to $M_*\sim10^{10}-10^{12}\,M_{\odot}$ and excluding LINERs as well as composite galaxies, as done by \citet{2024MNRAS.532.2320G}, our BPT-\textsc{Seyfert} fractions increase to 36.5\% for strongly barred, 26.8\% for weakly barred galaxies, and 24.8\% for unbarred galaxies, broadly similar to their reported values of 31.6\%, 23.3\%, and 14.2\%, respectively.
These results suggest that strong bars can efficiently supply fuel to the nuclear regions, thereby triggering or enhancing AGN activity.

However, in the sample with low mass and low SFR, the fraction of powerful AGN in unbarred galaxies is comparable to or even slightly higher than that in barred systems. This may reflect different fueling modes in gas-poor galaxies. In such systems, unbarred galaxies may sustain AGN activity through secular or stochastic accretion \citep[][]{2019MNRAS.484.4360A}, while barred AGN hosts may deplete their global gas reservoirs earlier due to efficient bar-driven inflows. As a result, barred galaxies lose their advantage in nuclear fueling and appear to host fewer powerful AGN in a statistical sense. However, this effect appears less prominent in massive barred galaxies, which could be related to the more efficient gas transport capabilities of strong bars. As shown in Figure~\ref{GR_M}, massive galaxies are more likely to host strong bars, these robust bars exert significantly stronger non-axisymmetric gravitational torques \citep[][]{2003MNRAS.341.1179A}, which can efficiently channel even sparse residual gas toward the center, maintaining nuclear activity despite low global gas fractions. This likely suggests that bars act as a primary catalyst for nuclear activity during a specific evolutionary window \citep[][]{2020MNRAS.494.5839K}.

We further note that the AGN fraction declines in the oldest galaxies ($D_n4000\gtrsim1.9$), and the differences in both AGN incidence and powerful AGN fraction between different samples largely vanish, implying that bar-driven fueling is effective primarily in gas-rich systems, but becomes inefficient once the cold gas is exhausted.

Moreover, we find that the dependence of AGN incidence on bar length is non-monotonic. In the strong bar sample, the AGN and BPT-\textsc{Seyfert} fractions increase with bar length at small to intermediate scales ($R_\mathrm{bar} < 8$ kpc), while at larger bar sizes the fractions show little further increase or even decline, as shown in Figure~\ref{agnfrac_barpara}. This suggests that the probability of triggering AGN activity may peak at an intermediate stage of secular evolution. In the early phases, the influence on gas dynamics of developing bars is still ramping up and may not yet sustain peak accretion \citep[][]{2015MNRAS.454.3641F}. In more evolved systems, sustained bar-driven inflow episodes and central star formation may have partially depleted the cold gas reservoir or increased the central mass concentration, thereby stabilizing the disk against further accretion \citep[][]{2026arXiv260222875U}. As a result, AGN fueling is most effective when bar-driven transport is well developed and substantial gas remains available in the disk.

Interesting, we find no other strong dependence of AGN activity on normalized bar length or bar ellipticity, suggesting that AGN triggering appears to be more closely related to the presence of a bar rather than to its structural properties.
In agreement with our results, \citet{2022A&A...661A.105S} found that bar is an important mechanism for fueling AGN activity, but parameters such as bar size, ellipticity, and effective surface brightness are not correlated with AGN activity. Recent work further suggests that while bars contribute significantly to low–moderate luminosity AGN, the most luminous AGN likely require mergers \citep[][]{2026A&A...707A.152L}.
This supports a scenario in which bars drive gas to the central kpc, after which secondary mechanisms such as turbulence or nested bars transport gas inward \citep[][]{1989Natur.338...45S, 2023ApJ...958...77L(nestedbar)}, ultimately fueling AGN \citep[][]{1999ApJ...525..691S}. Notably, weakly barred galaxies exhibit BPT-\textsc{Seyfert} fractions that are comparable to, or even slightly lower than those of unbarred systems. This may indicate that while structural parameters of bar do not strongly modulate AGN incidence, extremely weak bars may be insufficient to significantly influence nuclear activity. These results suggest that bar is an important mechanism in fueling AGN, while other mechanisms can also play a major role \citep[][]{2022A&A...661A.105S, 2015MNRAS.448.3442G, 2024MNRAS.532.2320G}.

Altogether, these results reinforce the role of bars as efficient drivers of central fueling and secular black hole growth.

\subsection{Limitations and Prospects}

A major strength of this study is the large and statistically robust sample from DESI DR1 and cross-matched with Galaxy Zoo DESI. With over $10^5$ galaxies in total and $\sim3\times10^4$ well-classified disks, this dataset constitutes one of the largest bar surveys to date. The large sample size allows us to dissect trends as a function of bar strength, stellar mass, color, and SFR with fine resolution, and to detect relatively subtle statistical effects. 
Moreover, we use the machine-learning morphological classifications developed by \citet{2023MNRAS.526.4768W}, trained on Galaxy Zoo citizen science data. These classifications offer robust and homogeneous bar identifications across the sample and enable a consistent comparison between bars of different prominence.
The spectroscopic data from DESI also enable uniformly derived SFR, metallicities, and AGN indicators for all galaxies, ensuring consistent comparisons among barred and unbarred populations.

Despite these advantages, several limitations should be taken into account. Our sample is confined to the local universe ($z<0.05$) and relatively luminous galaxies, so the trends we observe may not extend to higher redshift or lower masses. The bar classifications, while robust in aggregate, depend on threshold choices and image quality. For example, very weak bars may be misclassified as unbarred by the GZD machine-learning-based classification. The strong/weak bar dichotomy is also inherently somewhat arbitrary, and it may be more appropriate to view bar strength as a continuum, as suggested by \citet{2023MNRAS.524.3166E}, quantifying bar strength enables investigating the more fundamental impact that bars have on their host galaxies. Although our results suggest a strong association between bar-driven inflows and black hole fueling, more detailed observations or hydrodynamical simulations will be required to verify the underlying mechanisms. 

This work points toward several promising directions for future investigation. Deeper and higher-resolution imaging from upcoming surveys, such as DESI data releases (DESI DR2), the Chinese Space Station Survey Telescope \citep[CSST,][]{2026SCPMA..6939501C}, the Euclid mission \citep{2025A&A...697A...1E}, and the Legacy Survey of Space and Time \citep[LSST,][]{2019ApJ...873..111I}, will enable measurements of bar fraction and properties in fainter and more distant galaxies, testing how these trends evolve over cosmic time. High-resolution surface photometry also provides a more robust quantification of bar strength.
Mapping gas flows and star formation across bars has been extensively explored using IFU surveys such as SDSS-IV MaNGA \citep[][]{2020MNRAS.499.1116F, 2024ApJ...973..129G, 2024ApJ...973..116D}, providing detailed insight into how bars drive central gas inflow and star formation. Upcoming spatially resolved spectroscopic surveys (e.g., SDSS-V Local Volume Mapper; \citealt{2024AJ....168..198D}) will further expand this work to larger samples and higher spatial resolution. Building on existing studies that have investigated the gas content of barred galaxies \citep[e.g.,][]{2012MNRAS.424.2180M, 2017ApJ...835...80C, 2021MNRAS.507.4389G}, expanding to more comprehensive multi-wavelength data (CO, HI, and infrared observations) will help quantify the gas reservoirs in barred and unbarred galaxies. This will provide a more complete picture of how bars regulate the gas supply across different galactic environments. With the growing wealth of imaging, spectroscopic, and multi-wavelength data from CSST and other surveys, future studies will continue to refine our understanding of how bars shape galaxy evolution.

\section{Summary}
\label{sec:summary}

We have analyzed a large sample of nearby galaxies drawn from DESI DR1 cross-matched with Galaxy Zoo DESI morphological classification and several DESI value-added catalogs. Our principal findings are summarized below:

\begin{enumerate}
  \item The final cross-matched sample contains 33,201 disk galaxies. Within disk sample, barred galaxies constitute $35.67\%$ (11,843 galaxies), with strongly barred and weakly barred galaxies making up $10.57\%$ (3,508 galaxies) and $25.10\%$ (8,335 galaxies), respectively.
  \item The color and stellar-mass distributions show that weak bars are relatively common among bluer, lower-mass systems, while strong bars are preferentially found in redder, more massive disks. The total bar fraction displays a bimodal pattern across the $(g-r)$--$M_\ast$ plane. On average, strongly barred galaxies are more metal-rich than unbarred ones.
  \item Bars enhance the global star formation budget within the specific mass range of $M_\ast \sim 10^{9.5}-10^{10.5}\,M_{\odot}$. However, a clear mass dependence emerges in central regions. Strongly barred galaxies exhibit enhanced fiber SFRs within $M_\ast \sim 10^{9}-10^{9.5}\,M_{\odot}$, but show lower central sSFRs than unbarred systems at $M_\ast \sim 10^{9.5}-10^{10.5}\,M_{\odot}$. This supports a dual-role evolutionary sequence that bars initially catalyze central star formation but eventually promote long term quenching by accelerating gas consumption and facilitating bulge growth. This transition is further reflected in the non-monotonic dependence on bar length, where star formation peaks at $R_\mathrm{bar} \sim 8-12$ kpc. At $R_\mathrm{bar} > 12$ kpc, the star formation activity declines, as these long bars have typically undergone a more extended period of growth and correspond to more evolved systems, where bars have had sufficient time to redistribute and deplete the gas reservoirs, eventually leading to a reduced capability for the bar to significantly boost star formation.
  \item Our analysis indicates that barred disks exhibit higher AGN and BPT-\textsc{Seyfert} fractions, with strongly barred galaxies showing higher AGN incidence than weakly barred or unbarred systems. Examining the [OIII] $\lambda5007$ luminosity of AGN hosts, we find that the fraction of powerful AGN is also higher in strongly barred galaxies, except in systems with low stellar mass or low level SFR. These results indicate that bars play an important mechanism for fueling AGN, and are strongly associated with AGN activity. However, bar structural parameters such as length and ellipticity show only weak correlation with AGN incidence, suggesting that secondary mechanisms such as turbulence or nested bars are also necessary in AGN triggering.
\end{enumerate}

Taken together, our results support a picture in which bars act as engines of secular evolution: by funneling gas inward, they promote circumnuclear star formation and can enhance AGN activity at least in certain host regimes, while the cumulative effect of bar-driven star formation contributes to gas depletion and the subsequent migration of barred systems toward redder colors and quiescence. Future high-resolution and spatially resolved observations will be valuable for exploring the detailed inflow–fueling pathways and the temporal sequence connecting bar-driven central activity and long-term quenching.

\begin{acknowledgments}

We thank the anonymous referee and editor for their critical comments and instructive suggestions that significantly improved the quality of the paper.
This work is supported by the National Key Research and Development Program of China (Nos. 2025YFF0510604, 2025YFF0511000, 2022YFA1602902), the National Natural Science Foundation of China (No. 12373019), and the China Manned Space Project (Nos. CMS-CSST-2025-A08, CMS-CSST-2025-A19). This project is also supported by the Young Talent Fund of Institutional Center for Shared Technologies and Facilities of National Astronomical Observatories Astronomical Observatories, CAS.

This research used data obtained with the Dark Energy Spectroscopic Instrument (DESI). DESI construction and operations is managed by the Lawrence Berkeley National Laboratory. This material is based upon work supported by the U.S. Department of Energy, Office of Science, Office of High-Energy Physics, under Contract No. DE–AC02–05CH11231, and by the National Energy Research Scientific Computing Center, a DOE Office of Science User Facility under the same contract. Additional support for DESI was provided by the U.S. National Science Foundation (NSF), Division of Astronomical Sciences under Contract No. AST-0950945 to the NSF’s National Optical-Infrared Astronomy Research Laboratory; the Science and Technology Facilities Council of the United Kingdom; the Gordon and Betty Moore Foundation; the Heising-Simons Foundation; the French Alternative Energies and Atomic Energy Commission (CEA); the National Council of Humanities, Science and Technology of Mexico (CONAHCYT); the Ministry of Science and Innovation of Spain (MICINN), and by the DESI Member Institutions: www.desi.lbl.gov/collaborating-institutions. The DESI collaboration is honored to be permitted to conduct scientific research on I’oligam Du’ag (Kitt Peak), a mountain with particular significance to the Tohono O’odham Nation. Any opinions, findings, and conclusions or recommendations expressed in this material are those of the author(s) and do not necessarily reflect the views of the U.S. National Science Foundation, the U.S. Department of Energy, or any of the listed funding agencies.
\end{acknowledgments}

\bibliography{ref}{}
\bibliographystyle{aasjournalv7}

\end{document}